\documentclass[12pt]{article}

\usepackage{amssymb,amsmath,mathrsfs,epstopdf,slashed,color}

\usepackage{ifpdf}
\ifpdf
\usepackage{graphicx}
\usepackage{hyperref}
\usepackage{cite}
\else
\usepackage[dvipdfmx]{graphicx}
\usepackage[dvipdfmx]{hyperref}
\usepackage[numbers]{natbib}
\usepackage{hypernat}
\fi

\allowdisplaybreaks[1]

\makeatletter

\@addtoreset{equation}{section}
\makeatother

\newcommand{\vol}{\mathrm{vol}}

\DeclareMathOperator*{\re}{Re \,}
\DeclareMathOperator*{\im}{Im \,}

\title{\bf Population in multi K\"ahler model}
\author{Yoske Sumitomo, S.-H. Henry Tye}

\begin{document}

\begin{titlepage}

\setcounter{page}{0}
  
\begin{flushright}
 \small
 \normalsize
\end{flushright}

\vskip 3cm
\begin{center}

{\Large \bf A Stringy Mechanism for A Small Cosmological Constant \\
 - Multi-Moduli Cases -}  

\vskip 2cm
  
{\large Yoske Sumitomo${}^1$ and S.-H. Henry Tye${}^{1,2}$}
 
 \vskip 0.6cm

 ${}^1$ Institute for Advanced Study, Hong Kong University of Science and Technology, Hong Kong\\
 ${}^2$ Laboratory for Elementary-Particle Physics, Cornell University, Ithaca, NY 14853, USA

 \vskip 0.4cm

Email: \href{mailto:yoske@ust.hk, iastye@ust.hk}{yoske at ust.hk, iastye at ust.hk}

\vskip 1.0cm
  
\abstract{\normalsize
 Based on the properties of probability distributions of functions of random variables, we proposed earlier a simple stringy mechanism that prefers the meta-stable vacua with a small cosmological constant $\Lambda$. 
 As an illustration of this approach, we study in this paper particularly simple but non-trivial models of the K\"ahler uplift in the large volume flux compactification scenario in Type IIB string theory, where all parameters introduced in the model are treated either as fixed constants motivated by physics, or as random variables with some given uniform probability distributions.
 We determine the value $w_0$ of the superpotential $W_0$ at the supersymmetric minima, and  find that the resulting probability distribution $P(w_0)$ peaks at $w_0=0$; furthermore, this peaking behavior strengthens as the number of complex structure moduli increases.
The resulting probability distribution $P(\Lambda)$ for meta-stable vacua also peaks as $\Lambda \to 0$, for both positive and negative $\Lambda$.
 This peaking/divergent behavior of $P(\Lambda)$ strengthens as the number of  moduli increases.
 In some scenarios for $\Lambda > 0$, the likely value of $\Lambda$ decreases exponentially as the number of moduli increases.
 The light cosmological moduli issue accompanying a very small $\Lambda$ is also mentioned.
  }
  
\vspace{1cm}
\begin{flushleft}
 \today
\end{flushleft}
 
\end{center}
\end{titlepage}

\setcounter{page}{1}
\setcounter{footnote}{0}

\tableofcontents

\parskip=5pt

\section{Introduction}

Recent cosmological data strongly suggests that our universe has an exponentially small positive cosmological constant $\Lambda \sim 10^{-122}M_P^4$ \cite{Riess:1998cb,Perlmutter:1998np} (for more accurate recent data, see \cite{Komatsu:2010fb} and references therein). On the other hand, string theory has so many possible meta-stable vacuum solutions that it should have at least a solution with such a small $\Lambda$ \cite{Bousso:2000xa}. So it leaves open the question why nature picks a vacuum solution with such a very small $\Lambda $, in units of the Planck scale. In a previous paper \cite{Sumitomo:2012wa}, we propose a plausible reason why this may happen within the context of string theory. 

The basic idea is very simple. A typical compactification in string theory involves many moduli and fluxes (see the review \cite{Douglas:2006es}). The moduli and their dynamics describe the string theory landscape. Stabilizing them will lead to a set of values for these moduli, so $\Lambda$ of a meta-stable vacuum will be a function of the flux parameters.  Many if not most of the moduli can take multiple (some discrete) values since each stabilized compactification involves a set of quantized fluxes. As a result, many of these moduli will take values within a range that may include zero. If (and this is a big if) $\Lambda$ is a product of some of them, then the probability distribution $P(\Lambda)$ of $\Lambda$ will naturally peak (even mildly diverge) at $\Lambda=0$.
In actual models, the functional dependence of $\Lambda$ on the parameters in the model is much more complicated. 
Here we find the functional form of  $\Lambda$ in terms of the flux parameters  in a stringy scenario and show that the peaking property of $P(\Lambda)$ at $\Lambda=0$ is present. Furthermore, this peaking strengthens as the number of complex structure moduli increases.

In \cite{Sumitomo:2012wa}, we work out the case for a single K\"ahler modulus
in the K\"ahler uplifting region \cite{Balasubramanian:2004uy,Westphal:2006tn,Rummel:2011cd,deAlwis:2011dp} of the Large Volume Scenario \cite{Balasubramanian:2005zx} in Type IIB string theory.
There we treat the few parameters in the model as random variables and show that a modest suppression of $\Lambda$ is achieved: despite of the non-trivial functional dependence of $\Lambda$ on the parameters,
$P(\Lambda)$ is peaked (actually diverges) at $\Lambda=0$, 
as the parameters are treated as random variables with uniform (or similarly smooth) distributions.
In this single K\"ahler modulus model \cite{Sumitomo:2012wa}, we see that the suppression of  $\Lambda$ is present but very modest.
In this paper, we consider the multi-moduli cases to check the validity of the basic idea. Here our probability and statistical analysis is largely based on the study by Rummel and Westphal \cite{Rummel:2011cd} on a simplified yet non-trivial model in the Large Volume Scenario. Although the functional dependence of $\Lambda$ on the parameters is more non-trivial, we still find that the probability distribution $P(\Lambda)$ for meta-stable vacua becomes more peaked at $\Lambda=0$ (even mildly diverges) as the number of moduli increases.
Although the expectation value $\left< \Lambda \right>$ does not seem to drop quickly as the number of moduli increases,
in some scenarios,
the likely value of $\Lambda$ on the other hand decreases exponentially as the number of complex structure moduli increases.
This interesting behavior emerges in a number of scenarios in implementing the distributions of the flux parameters.

To be more specific, consider a Swiss-cheese type of Calabi-Yau three-fold with $h^{1,1}$ number of K\"ahler moduli and $h^{2,1}$ number of complex structure moduli (so the manifold $M$ has Euler number $\chi(M)=2(h^{1,1}-h^{2,1})$ and we are mainly interested in negative $\chi(M)$).
The simplified model of interest is given by, setting $M_P=1$,
\begin{equation}
 \begin{split}
  V =& e^{K} \left(K^{I \bar{J}} D_I W D_{\bar{J}} {\overline W} - 3\left|W \right|^2\right),\\
  K =& K_{\rm K} + K_{\rm d} + K_{\rm cs}= -2    \ln \left({\cal V} + {\hat{\xi} \over 2} \right) -    \ln \left(S+\bar{S} \right) - 
    \ln \left(-i \int {\overline \Omega} \wedge \Omega \right),\\
  {\cal V} \equiv& {\vol \over \alpha'^3 } = \gamma_1 (T_1 + \bar{T_1})^{3/2} - \sum_{k=2}^{h^{1,1}} \gamma_k (T_k + \bar{T}_k)^{3/2}, \quad 
  \hat{\xi} =  -\frac{\zeta(3)}{4\sqrt{2}(2\pi)^3} \chi(M) \left( S + \bar{S} \right)^{3/2}, \\
  W =&  W_0(U_i,S) + \sum^{h^{1,1}}_{k=1} A_k e^{-a_k T_k},
 \end{split}
 \label{LVS effective potential}
\end{equation}
where $\Omega$ is the homomorphic three-form.
In the superpotential $W$, the flux contribution to $W_0 (U_i,S)$ depends on the dilation $S$ and the $h^{2,1}$ complex structure moduli $U_i$ ($i=1,2,..., h^{2,1}$), while the non-perturbative terms for $h^{1,1}$ K\"ahler moduli $T_k$ ($k=1, 2, .., h^{1,1}$) are introduced in $W$ \cite{Kachru:2003aw}. The dependence of $A_i$ on $U_i, S$ are suppressed.
The model also includes the $\alpha'$-correction (the $\hat{\xi}$ term) to the K\"ahler potential \cite{Becker:2002nn}.
This model was originally proposed for the Large Volume Scenario \cite{Balasubramanian:2005zx} (see also \cite{Conlon:2005ki,Cicoli:2008va,Gray:2012jy}), and has been further analyzed in the search of de-Sitter vacua \cite{Balasubramanian:2004uy,Westphal:2006tn,Rummel:2011cd,deAlwis:2011dp}.

Since we like to study the behavior of the expectation value of $\Lambda$ when the number of complex structure moduli fields is large, we employ a simple model motivated by the orientifolded orbifolds of $T^6$ \cite{Lust:2005dy,Rummel:2011cd}, given by
\begin{equation}
 \begin{split}
  \label{w01}
  K_{\rm d+cs} =& -   \ln \left(S+\bar{S}\right) -     \sum_{i=1}^{h^{2,1}} \ln \left(U_i + \bar{U}_i \right),\\
   W_0(U_i,S) =&  c_1 +\sum_{i=1}^{h^{2,1}} b_i U_i - S \left(c_2 + \sum_{i=1}^{h^{2,1}} d_i U_i\right),
 \end{split}
\end{equation}
where $c_i, b_i$ and $d_i$ are (real) flux parameters that may be treated as random variables with smooth probability distributions that allow the zero values. Here we are interested in the physical $\Lambda$ (instead of, say, the bare $\Lambda$), so the model should include all appropriate non-perturbative effects, $\alpha'$ corrections as well as radiative corrections.
We see that the above simplied model (\ref{LVS effective potential}) includes non-perturbative $A_k$ terms to stabilize the K\"ahler moduli and the $\alpha'$ correction $\hat \xi$ term to lift the solution to de-Sitter space. In the same spirit, all  parameters in the model, in particular the coupling parameters $c_i, b_i$ and $d_i$ in $W_0$ (\ref{w01}), should be treated as physical parameters that have included all relevant corrections.

Following the analysis of \cite{Aazami:2005jf,Dean:2006wk,Dean2008,Borot:2010tr,Chen:2011ac}, we expect that the probability of an extremum at positive vacuum energy to be a classically stable solution will be Gaussianly suppressed  (see also an estimation at supersymmetric $AdS$ in SUGRA \cite{Bachlechner:2012at}).
So the existence of solutions will put constraints on the parameters and we take $s= {\rm Re} (S)=1/g_s > 1$ for weak coupling.
It turns out to be a good approximation to stabilize the $U_i$ and $S$ at a supersymmetric minimum ($D_SW_0 = D_{U_i}W_0=0$) before turning on the corrections for the stabilization of the K\"ahler moduli, which then breaks supersymmetry. At the supersymmetric minimum with their axionic components sitting at zero, $u_i= {\rm Re} (U_i) >0$ (required for $K_{\rm d+cs}$ (\ref{w01}) to stay real) and $s>1$ are determined and
 \begin{equation}
 \label{w02}
w_0 \equiv  W_{0}|_{\rm min} = -  \frac{2(c_1+ s c_2) \Pi_{i=1}^{h^{2,1}} (1- s r_i)}{\sum_{i=1}^{h^{2,1}} (1+ s r_i) \Pi_{j \ne i} (1-sr_j) },
  \end{equation}
 where $r_i=d_i/b_i$ and $s$ is given as a function of the real random parameters $c_1,c_2, r_i$.

\begin{figure}[t]
   \begin{center}
    \includegraphics[width=13.5em]{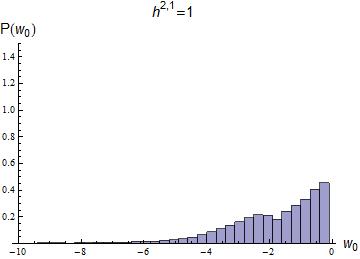}
    \includegraphics[width=13.5em]{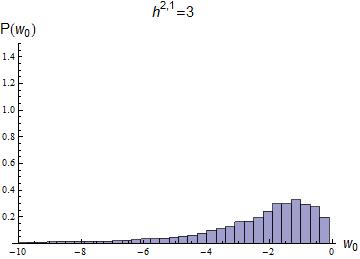}
    \includegraphics[width=13.5em]{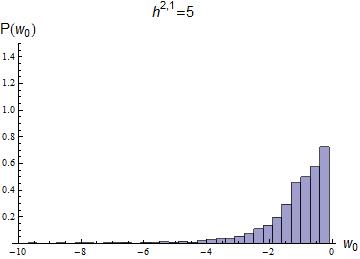}
    \includegraphics[width=13.5em]{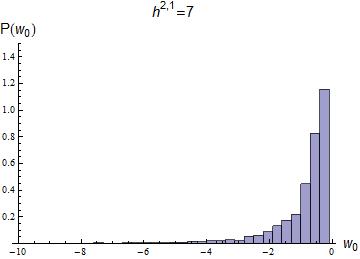}
    \includegraphics[width=13.5em]{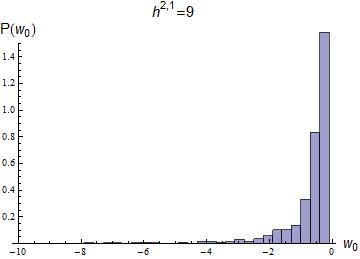}
    \includegraphics[width=13.5em]{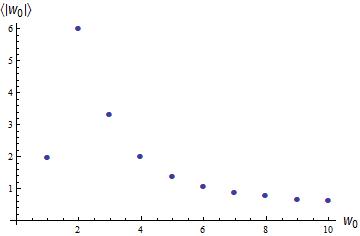}
   \end{center}
   \caption{\footnotesize The probability distribution $P(w_0)$ for $w_0$ (\ref{w02}) with $h^{2,1} = 1, 3, 5, 7, 9$ number of complex structure moduli respectively, where $w_0$ is the value of the superpotential $W_0$ at the supersymmetric solution.
   Here we present $P(w_0)$ in the range $-10 \leq w_0 \leq 0$ ($\int_{-10}^0  P(w_0) dw_0 =1$)  after solving for the complex structure moduli and the dilation at the supersymmetric point satisfying $s>1$ for weak string coupling and $u_i>0$ for the K\"ahler potential to be real. After some transient behaviors at small $h^{2,1}$, the peaking (at $w_0=0$) of $P(w_0)$ strengthens as $h^{2,1}$ increases. In the last figure, we show the behavior of the  expectation value $\left< |w_0 |\right>$.}
   \label{figw_0}
  \end{figure}

In general, the parameters will be fixed constants within the supergravity framework. However, it is the flux compactification property in string theory that allows us to compare solutions with different choices of values for the parameters.
Since the quantized fluxes of the higher-form field-strengths are expected to vary over large ranges of discrete values \cite{Bousso:2000xa,Denef:2004ze,Denef:2004cf}, the flux parameters $c_j, b_i$ and $d_i$ are expected to sweep through some smooth ranges of discrete values.
So we are justified to treat the parameters as variables with some suitable probability distributions and study the consequences. In this sense, the mechanism we suggest here can be considered as a stringy mechanism.
As we sweep through the (e.g., uniform) distributions for the flux parameters $c_1,c_2, b_i, d_i$, we see that each factor in the numerator of (\ref{w02}) easily passes through zero. So we expect the probability distribution $P(w_0)$ of  the value for the superpotential $w_0=W_{0}|_{\rm min}$ to become more peaked at $w_0 \to 0$ as the number $h^{2,1}$ of complex structure moduli increases (e.g., to hundreds). This behavior is illustrated in Figure \ref{figw_0}.

\begin{figure}[t]
   \begin{center}
    \includegraphics[width=13.5em]{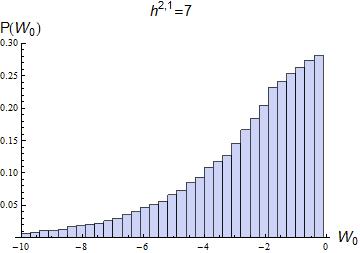}
    \includegraphics[width=13.5em]{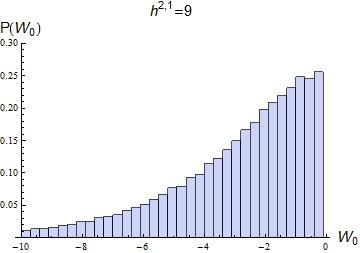}
    \includegraphics[width=13.5em]{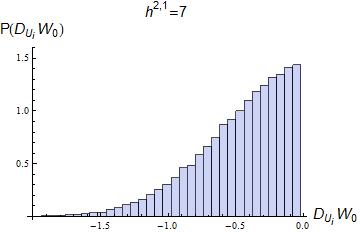}
   \end{center}
   \caption{\footnotesize The probability distribution $P(W_0)$ of $W_0$ (\ref{w01}) when the real components of the dilation and the complex structure moduli are also treated as random variables. Here we assume uniform distributions with ranges $-1\leq c_i, b_i, d_i \leq 1, \, 1<s \leq 5$, and $0< u_i \leq 1$. The first 2 figures are for $P(W_0)$ for $h^{2,1}=7, 9$. The third figure is for 
  $P(D_{U_i} W_0)$ of $D_{U_i} W_0$ for $h^{2,1}=7$. 
  }
   \label{randomW0}
  \end{figure}

It is interesting to compare $P(w_0)$ of $w_0$ (\ref{w02}) after stabilization with the probability distribution $P(W_0)$ of $W_0$ (\ref{w01}) before stabilization.  Let us choose the same uniform distributions for $c_i, b_i, d_i$ used in Figure \ref{figw_0}, but instead of solving for $u_i$ and $s$, we treat them as random variables with uniform distributions with range $0 \le u_i \le 1$ and $1 < s \leq 5$ (while the axionic modes are suppressed). With these uniform distributions,  we show $P(W_0)$ in Figure \ref{randomW0}; it is clearly smooth at $W_0=0$. We note that the probability distributions $P(D_{U_i} W_0)$ for the variables $D_{U_i} W_0$ are also smooth with little or no preference for the zero value, as expected. If $P(W_0)$ and $P(D_{U_i} W_0)$ are truly independent distributions, then fixing $D_{U_i} W_0=0$ (and $D_SW_0=0$) will not change the distribution $P(W_0)$; that is, $P(w_0)$ should be the same as $P(W_0)$. However, this is clearly not the case when we compare Figure \ref{figw_0} (for $P(w_0)$) and Figure \ref{randomW0} (for $P(W_0)$). This is because the two ``random'' distributions $P(W_0)$ and $P(D_{U_i} W_0)$ are actually correlated. In general, given any specific model, we expect correlations between these two distributions (as well as those for $DDW_0$ and $DDDW_0$). It is this correlation that leads to the peaking feature in $P(w_0)$ in Figure \ref{figw_0}.

\begin{figure}[t]
   \begin{center}
    \includegraphics[width=13.5em]{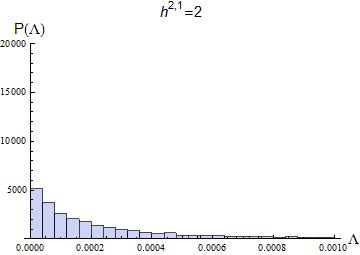}
    \includegraphics[width=13.5em]{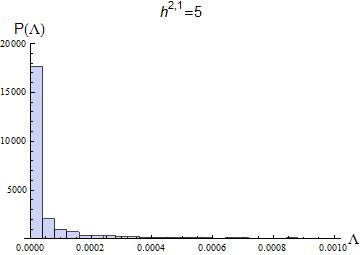}
    \includegraphics[width=13.5em]{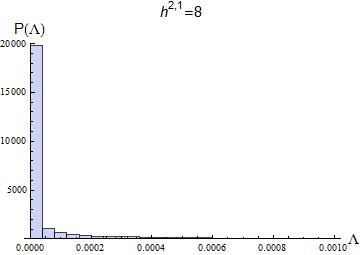}
    \end{center}
   \caption{\footnotesize The probability distribution $P(\Lambda)$ of the cosmological constant $\Lambda$ at meta-stable vacua as a function of $h^{2,1}= 2, 5, 8$ number of complex structure moduli and a single K\"ahler modulus ($h^{1,1} =1$). 
 Although the range is $0 \le \Lambda \lesssim 1$, the probability distributions for only $0 \leq  \Lambda \leq 10^{-3}$ are shown. $P(\Lambda)$ becomes more peaked at $\Lambda=0$ as $h^{2,1}$ increases.
       }
   \label{figLambda}
  \end{figure}

 Next we insert the value $w_0$ into the superpotential $W$ which is then inserted into the potential $V$. This allows us to find $\Lambda$ for the de-Sitter meta-stable vacua via the stabilization of the K\"ahler moduli.
 This hierarchical setup effectively reduces the number of moduli fields when we reach the energy level for the K\"ahler moduli stabilization with determined $s$ and the complex structure moduli $u_i$, and thus helps to enhance the probability for $dS$ vacua.  Since the SUGRA approximation is valid only if the scale of the potential $V$ (as measured for example  by the barrier height $V_H$) is around or below the Planck scale, we have to restrict ourselves to such a valid set of potentials $V$ in the determination of $\Lambda$. Typically, this will restrict us to $\Lambda \le 1$. 
Among these meta-stable vacua, we expect that the probability distribution $P(\Lambda)$ for $\Lambda$ will also peak at $\Lambda=0$. The numerical result on $P(\Lambda)$ for the single K\"ahler modulus scenario is shown in Figure \ref{figLambda}.

Although the peaking property of $P(\Lambda)$ is essential for the preference of a very small $\Lambda$, it is not enough. There are more than one way to implement the (uniform) distributions for the random parameters. In a number of scenarios of distributions of the random parameters, 
we find that the expectation value $\left< \Lambda \right>$ for de-Sitter vacua more or less stays constant (or changes little) as the number of complex structure moduli increases. That is, $P(\Lambda)$ in these scenarios has a long tail outside the $\Lambda \sim 0$ region. 

 \begin{figure}[t]
   \begin{center}
    \includegraphics[width=20em]{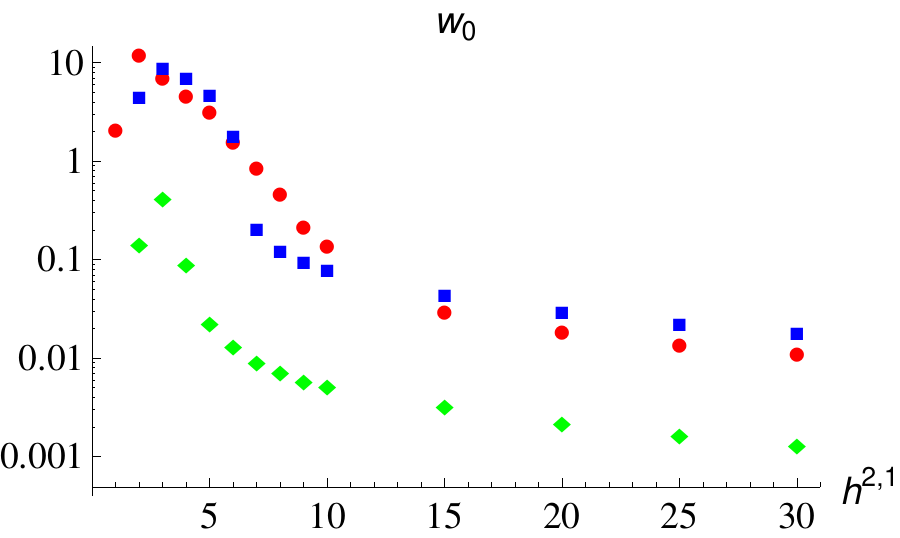}
    \includegraphics[width=20em]{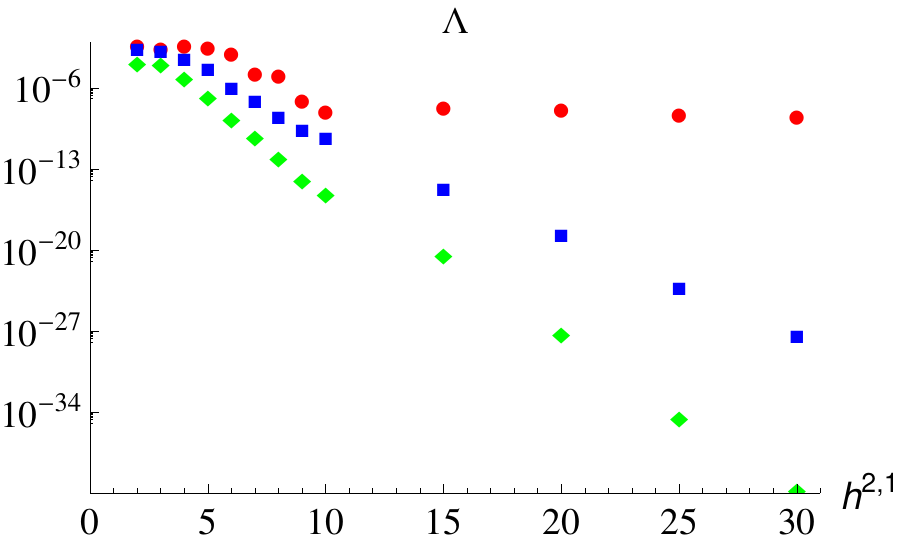}
   \end{center}
   \caption{\footnotesize  The LHS shows the average $\left< |w_0| \right>$ (red circles), $|w_{0}|^{80\%}$ (blue squares) and $|w_{0}|^{10\%}$ (green diamonds) as a function of the number $h^{2,1}$ of complex structure moduli. The RHS shows $\left< \Lambda \right>$ (red circles), $\Lambda^{80\%}$ (blue squares)  and $\Lambda^{10\%}$ (green diamonds) as a function of $h^{2,1}$. That is, there is  a $Y\%$ probability that $\Lambda$ of the meta-stable vacua will fall in the range $\Lambda^{Y\%} \ge \Lambda \ge 0$. For example, $\Lambda^{10\%} \simeq 4.8 \times 10^{-28}$ at $h^{2,1}=20$ and $\simeq 1.5 \times 10^{-41}$ at $h^{2,1}=30$ (green diamonds).
The values are for the case where the $b_i$ parameters are fixed. For comparison, we also show on the LHS  the average $\left< w_0 \right>$ and the corresponding $w_{0}^{80\%}$ and $w_{0}^{10\%}$. 
}
   \label{values-for-peaking}
  \end{figure}

The result of one specific scenario in shown in Figure \ref{values-for-peaking}. Here we find that the drop of $\left< \Lambda \right>$ slows down appreciably as  $h^{2,1}>10$, even though $P(\Lambda)$ becomes more peaked at $\Lambda \to 0$.
To get a better feeling of the peaking property of $P(\Lambda)$ for $\Lambda \ge 0$, let us introduce $\Lambda^{Y\%}$, defined by
$\int_0^{\Lambda^{Y\%}} \, P(\Lambda) \, d\Lambda = Y\%$. That is, there is  a $Y\%$ probability that $\Lambda$ of the meta-stable vacua will fall in the range $\Lambda^{Y\%} \ge \Lambda \ge 0$. (Note that $\Lambda^{100\%} \simeq 1$.) In Figure \ref{values-for-peaking},  we show $\left< \Lambda \right>$, $\Lambda^{80\%}$ and $\Lambda^{10\%}$ as a function of the number of complex structure moduli. For example, at $h^{2,1}=20$, while $\left< \Lambda \right> \simeq 1.2 \times 10^{-8}$,
$\Lambda^{80\%} \simeq 2.0 \times 10^{-19}$ and $\Lambda^{10\%} \simeq 4.8 \times 10^{-28}$.
For $h^{2,1}=30$, we find that $\Lambda^{80\%} \simeq 3.6 \times 10^{-28}$ and $\Lambda^{10\%} \simeq 1.5 \times 10^{-41}$; that is, there is a $80 \%$ chance that $\Lambda \lesssim 3.6 \times 10^{-28}$  and a $10 \%$ chance that $\Lambda \lesssim 1.5 \times 10^{-41}$. So we see that there is a reasonable probability that the likely $\Lambda$ drops exponentially with respect to the number of complex structure moduli.

 \begin{figure}[t]
   \begin{center}
    \includegraphics[width=13.5em]{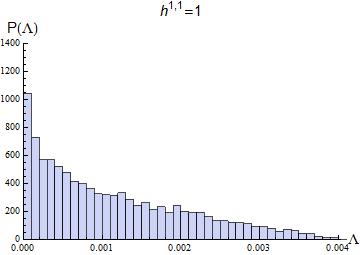}
    \includegraphics[width=13.5em]{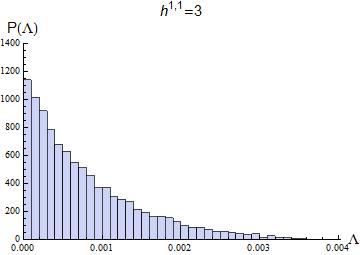}
    \includegraphics[width=13.5em]{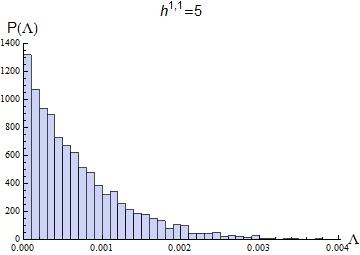}
   \end{center}
   \caption{\footnotesize  The probability distribution $P(\Lambda)$ for the $h^{1,1} = 1, 3, 5$ K\"ahler moduli cases with uniformly distributed $W_0, A_i$. Note that $P(\Lambda=0)$ is increasing (slowly) as $h^{1,1}$ increases.
}
   \label{fig3K}
  \end{figure}

  We also consider the multi-K\"ahler moduli cases. In Figure \ref{fig3K}, we show $P(\Lambda)$ for $h^{1,1}=1, 3, 5$ K\"ahler moduli cases (where the complex structure moduli and dilation are suppressed). We see that the peaking behavior also strengthens as the number of K\"ahler moduli increases.
For positive $\Lambda$, we have
$ \left< {\Lambda}\right> \sim 0.00111 \left(h^{1,1}\right)^{-0.282} e^{-0.0138\, h^{1,1}}$,
where uniformly distributed $W_0, A_i$ are assumed.
Here, the coefficient in the exponent is actually too small to be taken seriously.
However, if we have sharply peaked distributions in $W_0$ and $A_i$, a clear exponential suppression in $\left< \Lambda \right>$ as a function of $h^{1,1}$ is expected. We shall illustrate these features with some simplified models.

In short, we see that $P(\Lambda)$ becomes more peaked at $\Lambda=0$ as the number of moduli increases.
This feature is robust.
On the other hand, the behavior of $\left< \Lambda \right>$ (or $\sqrt{\left< \Lambda^2 \right>}$) is sensitive to the way we implement the  random parameter distributions and other inputs. The model studied here is chosen for its simplicity.
The numerical data presented here are for the specific scenarios described. There may be other interesting scenarios within this simple model.
To go beyond this simple model, it is important to examine more realistic models to check (1) whether the peaking of $P(\Lambda)$ at $\Lambda=0$ is as robust as we believe, and (2) what models will lead to an exponentially decreasing $\Lambda$ (either $\left< |\Lambda| \right>$ or $\Lambda^{Y\%}$) as a function of the number of moduli. A better understanding of the underlying physics should allow us to determine the value of $\Lambda$ as a function of the number of moduli. An exponentially decreasing $\Lambda$ may be used as a criteria to select particular regions of the landscape.

In general, the probability that an extremum happens to be a meta-stable vacuum is Gaussianly unlikely as the number of moduli increases, as discussed in \cite{Aazami:2005jf,Dean:2006wk,Dean2008,Borot:2010tr,Chen:2011ac,Bachlechner:2012at}.
It is our goal to find scenarios where an exponentially small $\Lambda$ for a meta-stable vacuum is preferred when the number of complex structure moduli is large, as one expects to be the case in any realistic model. 
In fact, some of the commonly studied Calabi-Yau 3-folds with one K\"ahler modulus are smooth hypersurfaces in ${\mathbb CP}^4$ with hundreds of complex structure moduli  (see e.g. \cite{Denef:2004dm}.
So an exponentially small $\Lambda \ge 0$ hopefully may be naturally realized in this framework.

 Note that having a very small $\Lambda$ does not necessarily solve the $\Lambda$ problem completely. In this setup, the expectation value of the mass (squared) of the lightest modulus tends to decrease quickly as $h^{2,1}$ increases.
So intuitively we may have the cosmological moduli problem \cite{Coughlan:1983ci,Banks:1993en,deCarlos:1993jw}. Since ways to avoid this problem within supergravity has been proposed and studied  \cite{Linde:1996cx,Takahashi:2010uw,Takahashi:2011as,Kallosh:2011qk},  it remains to be seen whether the light modulus mass is really a cosmological problem (within Type IIB string theory) as one may naively envision.

In summary, we find that the study of multi-moduli scenarios is a promising direction in the search for the reason why the observed $\Lambda$ is so small.
Tremendous amounts of effort have been spent on the search for a string vacuum with a standard model of  three families of quarks and leptons. We believe it is worthwhile to search for regions of vacua in the stringy cosmic landscape that have a chance of providing a naturally small positive $\Lambda$. Combining these two criteria in a search may be more effective than the searches carried out so far. On a separate note, we hope that this functional probability distribution idea can find applications in other phenomena.

The rest of the paper is organized as follows. Section \ref{sec:background-review-} reviews some simple probabilistic properties of functions of random variables and the single K\"ahler modulus model.
Section \ref{sec:multi-compl-struct} discusses the multi-complex structure moduli case. Here we find the supersymmetric solutions for the complex structure moduli and the dilation and then insert them into the single K\"ahler modulus model to solve for $\Lambda$.
Section \ref{sec:multi-khaler-moduli} discusses the multi- K\"ahler cases.
Here, we also consider  some simplified versions of the general multi-moduli cases.
Section \ref{sec:discussions-remarks-} presents the summary and discussions.
Some details are relegated to the appendices.

\section{Background and Review \label{sec:background-review-}}

We begin with a review of the basic probability distribution of a function of random variables when the distributions of the random variables are given. In particular, we emphasize on the peaking property in the resulting probability distribution, since this is relatively insensitive to the details. 
We explain how a product of random variables gets a sharply peaked (at zero) probability distribution.  We also point out the importance of the overall distributions to the expectation values. 
Then we review how this feature emerges for the probability distribution of $\Lambda$ in the single K\"ahler modulus model in a class of models in Type IIB string theory well studied already, basically following \cite{Sumitomo:2012wa}. We shall use this model to provide the framework to study the multiple complex structure moduli scenarios.
In appendix \ref{toy1}, we give some field theory toy models to illustrate some of the probabilistic properties that one may encounter in this analysis.

\subsection{Probability distribution of functions of random variables}

 Suppose we have $n$ random variables $y_i$ ($i=1,2, \cdots , n$), each with probability distribution $P_i(y_i)$, where $\int  P_i(y_i) dy_i =1$. Let us consider an arbitrary function $z=f(y_1, y_2, \cdots , y_n)$. Then the probability distribution $P(z)$ of $z$ is given by
 \begin{equation}
  \begin{split}
 P(z) = &\int dy_1dy_2 \cdots dy_n  P_1(y_1)  P_2(y_2) \cdots P_n(y_n) \, \delta (f(y_i) - z), \\ 
 \int  P(z) dz =& 1,
 \end{split}
 \label{normalization}
 \end{equation}
so the probability distribution $P(z)$ of $z$ can always be properly normalized, even when $P(z)$ diverges at $z=0$ and/or elsewhere.

Consider the following simple example:
  $z=y_1 y_2 \cdots y_n$ and each random parameter $y_i$ obeys the uniform distribution $P(y_i)=1/L$ with range $0\leq y_i \leq L$; then the probability distribution $P(z)$ of $z$ is given by (for $0 \le z \le L^n$)
  \begin{equation}
   \begin{split}
    P(z) = \int_{0}^L \frac{dy_1}{L} \frac{dy_2}{L} \cdots \frac{dy_n}{L} \, \delta (y_1 y_2 \cdots y_n - z) = {1\over (n-1)! L^n} \left( \ln {L^n\over |z|} \right)^{n-1}.
   \end{split}
   \label{product distribution}
  \end{equation}
  Therefore $P(z)$ of a product of random parameters has a mild divergent peak at $z=0$. This divergent behavior strengthens as $n$ increases. Actually this logarithmic divergent behavior of $P(z)$ at $z=0$ (\ref{product distribution}) is present for any $y_i$ distribution that smoothly includes $y_i=0$.

  Given the probability distribution for $y_i$, we can determine its expectation value.
  Here, for $z=y_1 y_2 \cdots y_n$, the expectation value of the $p$th moment $\left<z^p\right>$ of $z$ is given by 
  \begin{equation}
  \label{zexp1}
   \left< z^p \right> = \left< y^p_1 \right> \cdots \left< y^p_n \right>
  \end{equation}
  following from the property of Mellin integral transformation.
   Since $\left<y_i\right>=L/2$ for the uniform distribution, we have 
   \begin{equation}
   \left< z \right> = \left({L\over 2}\right)^{n}=  e^{-n \ln (2/L)  }.
    \label{eq:exponential suppression in product distribution}
   \end{equation}
 For $L<2$, we see that $ \left< z \right> $ decreases exponentially as $n$ increases;  so the divergent behavior at $z=0$ is correlated to the smallness of $\left< z \right>$. However, $ \left< z \right> $ increases exponentially if $L>2$. In this case, the long tail (at large $z$) of $P(z)$ drives $ \left< z \right> $. In short, we see that the behavior of $\left< z \right>$ (as a function of $n$) is very sensitive to the property of $P(z)$, especially at large $z$.

 To conclude, we like to mention that $P(z)$ is in general smooth at $z=0$ if $z$ is a sum of terms that involve independent random variables (see e.g. \cite{Sumitomo:2012wa} for more details). The problem now is to find the functional form of the cosmological constant $\Lambda$ of meta-stable vacua in terms of the random parameters entering the model and see whether its distribution $P(\Lambda)$ has the nice peaking property at zero and whether its expectation value or $\Lambda^{Y\%}$ decreases exponentially fast as a function of the number of moduli.

\subsection{The single K\"ahler modulus model \label{sec:revi-analyt-study}}

We briefly review the analytic study of the probability distribution $P( {\Lambda})$ 
in the single K\"ahler modulus scenario in type IIB string theory \cite{Sumitomo:2012wa}.
Although no potential is generated for the K\"ahler modulus at the classical tree level, non-perturbative terms can be introduced into the superpotential $W$ and $\alpha'$-corrections into the K\"ahler potential to break the no-scale structure, so that a non-trivial potential is generated at next leading order.
The tree level potential basically controls the stabilization of complex structure moduli and dilaton.
Here we simply assume the complex structure sector is stabilized at high-energy levels and therefore is believed to have been integrated out in the low energy effective model.
Later, we shall introduce the complex structure sector, solve it at the supersymmetric point and insert the solution into this model to solve for the stabilization of the K\"ahler modulus which then breaks supersymmetry.

The model of interest is given in \cite{Balasubramanian:2004uy,Westphal:2006tn,Rummel:2011cd,deAlwis:2011dp}, also known as {\it K\"ahler uplifting model} :
\begin{equation}
 \begin{split}
  K =& -2   \ln \left({\cal V} + {\hat{\xi} \over 2} \right), \quad 
  {\cal V} = \gamma_1 (T_1 + \bar{T_1})^{3/2}, 
\\
  W =& W_0  + A_1 e^{-a_1 T_1},
\end{split}
 \label{eq:potential for single Kahler}
\end{equation}
where the $\alpha'$-correction related to the $\hat{\xi}$ term \cite{Becker:2002nn} and non-perturbative correction with the parameter $A_1$ \cite{Kachru:2003aw} are given. The parameters $W_0, A_1$ will be treated as random variables.
We work within the approximation
\begin{equation}
 \begin{split}
  {\hat{\xi} \over {\cal V}} \ll 1, \quad
  \left| {A_1 e^{-a_1 T_1} \over W_0} \right| \ll 1,
 \end{split}
 \label{eq:perturbative approximation for single Kahler}
\end{equation}
to simplify the analytical study \cite{Rummel:2011cd}.
We shall see that this approximation is quite reasonable for the analysis of small $\Lambda$.
The first approximation is a sort of the large volume approximation which is powerful for suppressing the other unwanted stringy corrections.

\begin{figure}[t]
 \begin{center}
  \includegraphics[width=25em]{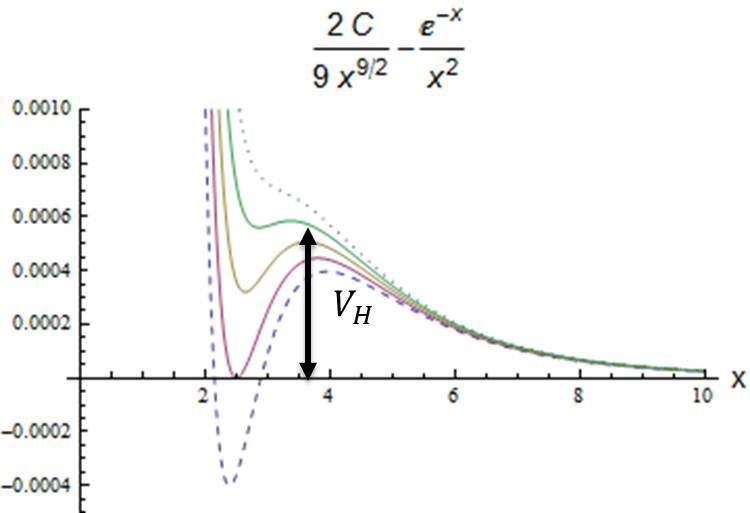}
 \end{center}
 \caption{\footnotesize The potential $V$ for different values of $C$. Only the bracket part in $V$ (\ref{eq:westphal potential}) is shown. The $V_H$ defined at $C \sim 3.89$ gives roughly the height of the potential barrier. }
 \label{fig:V_H}
\end{figure}

Since the imaginary part of the K\"ahler  modulus $T_1$ has a cosine type of potential, the extremal condition for this direction is satisfied when ${\rm Im} \, T_1 = 0$.
Therefore we focus only on real part $t_1 \equiv {\rm Re}\, T_1$.

Now the potential simplifies to \cite{Rummel:2011cd}
\begin{equation}
 \begin{split}
  {V  } \sim& {-W_0 a_1^3 A_1 \over 2 \gamma_1^2} \left({2C \over 9 x_1^{9/2}} - {e^{-x_1} \over x_1^2} \right),\\
  C = & {-27 W_0 \hat{\xi} a_1^{3/2} \over 64 \sqrt{2} \gamma_1 A_1}, \quad
  x_1 = a_1 t_1.
 \end{split}
 \label{eq:westphal potential}
\end{equation}
The stability condition $\partial_{x_1}^2 V >0$ at the extrema $\partial_{x_1} V =0$ with respect to $x_1$ is easy to analyze, and we get the parameter range for stable positive $\Lambda$:
\begin{equation}
C_0 \lesssim  C < C_1 \quad  \to \quad   3.65 \lesssim C < 3.89,
  \label{eq:stable region for single Kahler}
\end{equation}
where the lower bound is given by positivity of the minimum of $V$, while the upper bound is given by the stability constraint.
If we satisfy the combination of parameters $C$ inside this region, there is a stable solution in the range $2.50 \lesssim x_1 < 3.11$ at $\Lambda \ge 0$. Up to an overall factor, the potential $V$ (\ref{eq:westphal potential}) is shown in Figure \ref{fig:V_H}.

Our interest is to study the statistical property at small $\Lambda$.
In this approximation, the extremal condition gives us the linear relation between $x_1$ and $C$ by
\begin{equation}
 x_1 \sim \left({5\over 2}\right)^{-5/2} e^{5/2} C - 2.
\end{equation}
and we have
\begin{equation}
 \begin{split}
   {\Lambda} \equiv \left. {V  } \right|_{\rm min} \sim {1\over 9} \left({2 \over 5}\right)^{9/2} {-W_0 a_1^3 A_1 \over \gamma_1^2} \left(C - 3.65 \right).
  \label{1Kf}
 \end{split}
\end{equation}

Looking at $V$ (\ref{1Kf}), we note that the height of the potential barrier is relatively unchanged as $\Lambda$ varies from negative values to positive values, as shown in Figure \ref{fig:V_H}. So we may use the barrier height as a measure of the scale of the potential $V$,
\begin{equation}
 V_H =  {1\over 9} \left({2 \over 5}\right)^{9/2}{-W_0 a_1^3 A_1 \over \gamma_1^2} (0.24).
 \label{barrier}
\end{equation}
In general, the SUGRA approximation is valid if the scale of $V$ is less than the Planck (or string) scale. We shall require $V_H \le 1$ as the condition for the validity of our approximation.

In finding the probability distribution $P(\Lambda)$, the peaking behavior of the distribution comes essentially from this $V_H$. Since we would like to focus on the shape of distribution, we set the maximal value to be $V_H|_{\rm max} = 1$ here, such that the effective theory is safely applicable under the Planck scale.
For further simplification, we may set $a_1, \gamma_1, \hat{\xi} $ to constants and see the effect of two random variables $W_0, A_1$ on $P(\Lambda)$. (In \cite{Sumitomo:2012wa}, we also discuss cases where the other parameters are treated as random variables.)

\begin{figure}[t]
 \begin{center}
  \includegraphics[width=18em]{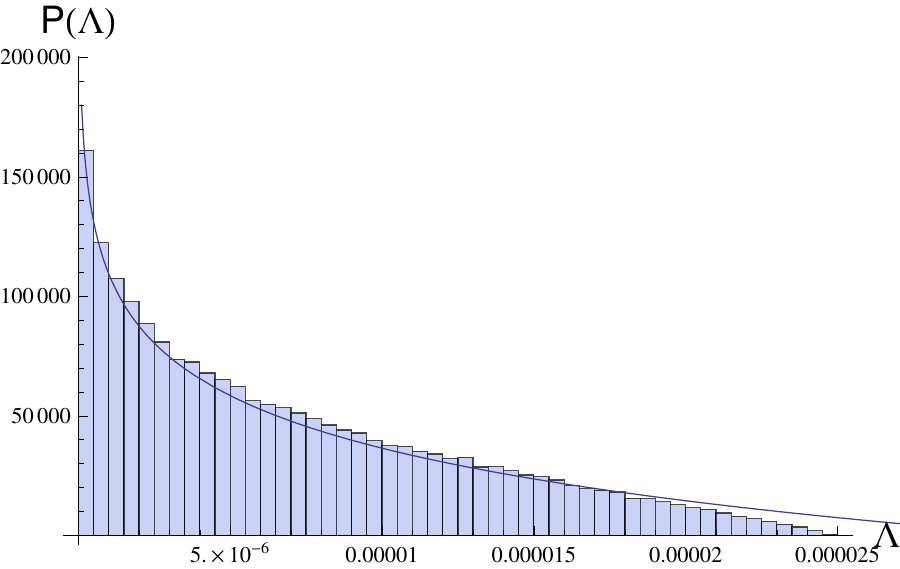}
  \includegraphics[width=18em]{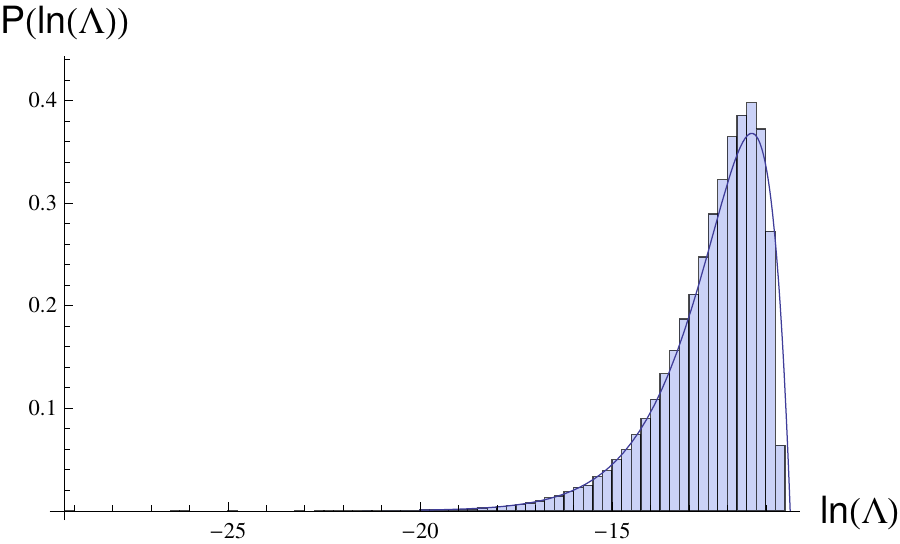}
 \end{center}
 \caption{\footnotesize The comparison between the analytic result (\ref{eq:PDF for Lambda}) (solid line) and the numerical result of the potential (\ref{eq:potential for single Kahler}) (histogram).
 Here we set $a_1=\gamma_1=  {\xi}=1$ and assume uniform distribution for parameters $W_0, A_1$.
 The plot in LHS shows the probability distribution $P(\Lambda)$ of $ {\Lambda}$, while the plot in RHS is  the probability distribution $P(\ln  {\Lambda})$ vs $\ln  {\Lambda}$.
 They are related via $P( {\Lambda}) \, d  {\Lambda} = P(\ln  {\Lambda}) \, d \ln  {\Lambda}$.}
 \label{fig:comparison-1K}
\end{figure}

If we simply assume $0\leq -W_0, A_1 \leq 1$ obeying uniform distribution, the probability distribution can be obtained (the detail is relegated to appendix \ref{sec:prob-distr-plambda}),
\begin{equation}
 P( {\Lambda}) = {2500 \sqrt{5} \over 3 \delta C} \ln \left[{3 \over 2500 \sqrt{5}} {\delta C \over  {\Lambda}  }\right].
  \label{eq:PDF for Lambda}
\end{equation}
where $\delta C = (C_1 -C_0)/C_1 = 0.0617$.
So we clearly see the divergent behavior as $\Lambda \to 0$. 
The stabilization dynamics is responsible for the divergent behavior in $P( {\Lambda})$. 
As we state in the introduction, the peaking behavior of $P( {\Lambda})$ is robust. 
Note that this divergent behavior is present for negative $\Lambda$ as well. For $\Lambda <0$, there is another branch of supersymmetric $AdS$ solutions; so $P(\Lambda)$ for $\Lambda <0$ is not a reflection of the $P( {\Lambda})$ shown above.

Let us compare $P(\Lambda)$ obtained analytically in the approximate potential (\ref{1Kf}) with that obtained numerically
for the potential (\ref{eq:potential for single Kahler}), as shown in Figure \ref{fig:comparison-1K}.
From the two figures there, we see their nice agreement with each other, especially at small $\Lambda$.

\section{Multiple complex structure moduli cases \label{sec:multi-compl-struct}}

So far we have reviewed the case with a single K\"ahler modulus assuming moduli stabilization of the complex structure and the dilaton at higher energy scale.   In the actual models, we have the dilaton $S$ and many complex structure moduli $U_i$ ($i=1,2,...,h^{2,1}=N_{\rm cs}$) which are assumed to be stabilized at a higher scale with fluxes as sources.
Here, we analyze in more detail their stabilization and see how it affects the probability distribution of $W_0$.
In \cite{Denef:2004ze}, a random distribution for the superpotential (and its derivatives) is assumed to investigate the entire landscape of SUGRA. Here, we probe parts of the SUGRA landscape in greater details.

Throughout this section, we focus on models of the form:
 \begin{equation}
  \begin{split}
    K_{\rm cs+d} =&-   \ln \left(S+\bar{S}\right) -     \sum_{i=1}^{h^{2,1}} \ln \left(U_i + \bar{U}_i \right),\\
   W_0 =& c_1 + \sum_{i=1}^{h^{2,1}} b_i U_i - \left(c_2 + \sum_{i=1}^{h^{2,1}} d_j U_j \right)S, 
  \end{split}
  \label{simplest model for complex}
 \end{equation}
 where $c_i, b_i, d_i$ are parameters from fluxes.
 This model is motivated by the orientifolded orbifolds of $T^6$ \cite{Lust:2005dy,Rummel:2011cd}, and is used for the purpose to see the tendency of the probability distribution of $W_0$ as a function of $h^{2,1}$ here.
Supersymmetry breaking comes from the interplay between the $\alpha '$ correction $\hat \xi$ term and the non-perturbative term reviewed in section \ref{sec:revi-analyt-study}. So it is a good approximation to solve for $U_i$ and $S$ at the supersymmetric point and then insert their values into the full system to solve for the stabilization of the K\"ahler moduli. 
It is worth noting that we basically require $\hat{\xi} >0$ (\ref{LVS effective potential}) for the stability analyses of K\"ahler moduli later, meaning that the number of complex structure modulii is greater than the number of K\"ahler moduli: $\chi(M)=2(h^{1,1}-h^{2,1})<0$.
As pointed out in  \cite{Rummel:2011cd}, this approximation is valid when the volume is large, as we always assume, and when $h^{2,1} \gg 1$, which is precisely the region we are interested in.

\subsection{Supersymmetric Minimum}

 Following the approach in  \cite{Rummel:2011cd}, we stabilize the dilation $S=s+ia$ and complex structure moduli $U_j=u_j+i{\nu}_j$ at the SUSY point to find $W_0|_{\rm min}$ and then insert it into the superpotential $W$ for the multi-K\"ahler moduli stabilization where SUSY will be broken due to the non-perturbative effects.
To find the supersymmetric values for $U_i$ and $S$, we have $D_SW_0=\partial_{S}W_0 +\partial_S K  W_0=0$ and $D_{U_i}W_0=0$, or 
\begin{align}
 \label{SUSY1}
  &{\rm Re} \left(D_SW_0\right)={\rm Re}\left(D_{U_i}W_0\right)=0,\\
 \label{SUSY2}
  &{\rm Im}\left(D_SW_0\right)={\rm Im}\left(D_{U_i}W_0\right)=0.
\end{align}
For $s>1$, (\ref{SUSY1}), (\ref{SUSY2}) have solutions so the imaginary modes decouple in (\ref{SUSY1}).
 So it is most convenient to work with the $a=\nu_i=0$ solution for the imaginary modes and
 focus on the real parts of the moduli fields in the region $s>1$ (for weak coupling) and $u_i>0$ (to ensure that $K_{\rm cs+d}$ stays real).
 Now $W_0$ becomes (defining $N_{\rm cs} \equiv h^{2,1}$ for convenience)
  \begin{equation}
  \label{mc1}
    W_0= (c_1- s c_2) + \sum_{i=1}^{N_{\rm cs}} (b_i -s d_i)u_i 
  \end{equation}
and $D_SW_0=0$ and $D_{U_i}W_0=0$ yield (assuming $s \ne 0$ and $u_i \ne 0$)
  \begin{align}
   \label{mc2}
    &(c_1 + s c_2) + \sum_{i=1}^{N_{\rm cs}} (b_i +s d_i)u_i=0 ,\\
   \label{mc3}
    &(c_1 - s c_2) - (b_i -s d_i)u_i + \sum_{j \ne i}^{N_{\rm cs}} (b_j -s d_j)u_j  =0.
  \end{align}

  Now ${N_{\rm cs}}=0$ and ${N_{\rm cs}}=1$ are special cases and are easy to solve. So let ${N_{\rm cs}} \ge 2$.
 We shall see that ${N_{\rm cs}}=2$ is also special.
 (\ref{mc3}) immediately gives
 \begin{equation}
 \label{mc4}
    v\equiv (b_1 -s d_1)u_1 = (b_2 -s d_2)u_2= \cdots   = (b_{N_{\rm cs}} -s d_{N_{\rm cs}})u_{N_{\rm cs}}.
  \end{equation}
So $u_i$ are solved in terms of $s$ and one of them, say $u_1$, or equivalently, $v$.
Then (\ref{mc3}) becomes
 \begin{equation}
 \label{mc5}
    (c_1 - s c_2) + (N_{\rm cs}-2) v = 0,
  \end{equation}
    and thus
    \begin{equation}
   \label{mc6}
    W_0|_{\rm, min}=  -2 {c_1 - s c_2\over N_{\rm cs}-2} =2v,
  \end{equation}
  where the second form is also applicable to the ${N_{\rm cs}} \ne 0$ cases.
  We see that $s=c_1/c_2$ for ${N_{\rm cs}}=2$.

  Now we have two equations (\ref{mc2}) and (\ref{mc5}) for two unknowns, $v$ (or $u_1$) and $s$, and we have
  for ${N_{\rm cs}} \ne 2$,
   \begin{equation}
 \label{mc7}
   (N_{\rm cs}-2) \frac{c_1 + s c_2}{c_1 - s c_2} = \sum_{i=1}^{N_{\rm cs}} \frac{b_i+s d_i}{b_i -s d_i}= \sum_{i=1}^{N_{\rm cs}} \frac{1+s r_i}{1 -s r_i}.
  \end{equation}
  where $r_i=d_i/b_i$.
 In general, we cannot solve analytically for $s$. However, once $s$ is determined by  (\ref{mc7}), we can express $W_ 0|_{\rm min} ({N_{\rm cs}})$ in terms of $s$,
   \begin{equation}
 \label{mc7b}
w_0\equiv  W_0|_{\rm min} ({N_{\rm cs}})=
     -  \frac{2(c_1+ s c_2)}{\sum_{i=1}^{N_{\rm cs}} {(1+s r_i)}/{(1 -s r_i)}}=
 -  \frac{2(c_1+ s c_2) \Pi_{i=1}^{N_{\rm cs}} (b_ i- s d_i)}{\sum_{i=1}^{N_{\rm cs}} (b_i+ s d_i) \Pi_{j \ne i} (b_j-sd_j) }.
  \end{equation}
  This is the formula quoted in the Introduction. 
  Recall that we like $s$ to be large and positive, say $s \approx {\cal O} (10)$.
 For the $N_{\rm cs}=0,1,2$ cases:
 \begin{equation}
  \begin{split}
   W_0|_{\rm min}({N_{\rm cs}}=0)=& 2 c_1, \quad s=-{c_1 \over c_2},\\
   W_0|_{\rm min}({N_{\rm cs}}=1) =& - 2 \frac{(c_1+sc_2)(b_1-sd_1)}{b_1+sd_1}, \quad s=\sqrt{c_1b_1\over c_2d_1},\\
   W_0|_{\rm min} ({N_{\rm cs}}=2) =& - \frac{(c_1+sc_2)(b_1-sd_1)(b_2-sd_2)}{b_1b_2-s^2d_1 d_2},  \quad s={c_1 \over c_2}.
  \end{split}
 \end{equation}
 In appendix \ref{sec:discussion-solving-s}, we further discuss the solution for $s$ in special cases.

 Treating $c_j$, $b_j$ and $d_j$ as random parameters, we allow them to take values such that each solution satisfies $s>1$ and $u_i >0$.
 We see that the each factor $(b_j-sd_j)$ in the numerator in (\ref{mc7b}) is allowed to pass through zero.
We will check numerically that there is actually a peaking behavior as the number of complex structure moduli increases.

Before introducing the next-leading order terms (both non-perturbative and the $\hat \xi$ term) to generate a potential for the K\"ahler moduli, the model has a no-scale structure so $|D_{T_k}W|^2=3|W|^2$. Since $V= e^{K} \sum_i |D_i W|^2$ (with $i$ running over complex structure moduli and dilaton) is semi-positive definite (with the supersymmetric minimum at $V=0$), the Hessian (mass squared matrix) is semi-positively definite, owing to the no-scale structure.
On the other hand, some of masses for the axionic modes turn out to be zero as discussed in \cite{Rummel:2011cd}.
Presumably, the non-perturbative corrections will generate cosine-like potentials, thus stabilizing the axionic fields.

The above values for $s$ and $u_i$ are solved at the supersymmetric point. Their values will be shifted after supersymmetry breaking \cite{Rummel:2011cd}.
The corrections are estimated of order
\begin{equation}
 {\delta (s, u_i) \over s, u_i} \sim {\cal O} \left({\hat{\xi} \over {\cal V}}\right).
\end{equation}
As we reviewed in section \ref{sec:revi-analyt-study}, the expansion of ${\cal O} ( \hat{\xi} / {\cal V} )$ is compatible with the analysis for smaller $\Lambda$, and thus we can neglect the small correction to the complex structure sector in our analysis.
It is worth commenting about the situation that total summed over corrections with large $N_{\rm cs}$ is not negligible.
In this case, the extremal condition in the perturbation $\partial_{\phi_i} \partial_{\phi_j} V \cdot \delta \phi_j = - \partial_{\phi_i} \delta V$ suggests roughly
\begin{equation}
 {\delta (s, u_i)} \propto {\cal O} \left({1\over N_{\rm cs}}\right).
\end{equation}
Therefore the correction may be suppressed by large $N_{\rm cs}$ instead.

\subsection{Some properties of $w_0=W_0|_{\rm min}$}

 We see that $w_0$ (\ref{mc7b}) together with the solution for $s$ (\ref{mc7}) have some interesting properties. 

 \begin{itemize}
  
\item Since $c_1, c_2, b_j, d_j$ are random variables, so, for any fixed $j$, $(b_j-sd_j) \to 0$ is surely allowed.  (\ref{mc7}) for $s$ indicates that when $(b_j-sd_j) \to 0$  for any $j$, the RHS blows up (in the absence of delicate cancellations) so LHS requires $s \to c_1/c_2$.  Now, in the vicinity of $b_j-sd_j= 0$, $s$ becomes independent of the other values of $(b_i, d_i)$ for $i \ne j$. 

\item  The factors $(b_i-sd_i)$ appear in the numerator of $w_0$ (\ref{mc7b}). Knowing that $s \to c_1/c_2$ is insensitive to $(b_i, d_i)$ for $i \ne j$ in the vicinity of $(b_j-sd_j)=0$, we can work out the probability distribution $P(z)$ for $z=b_i-sd_i$  for $i \ne j$,
 \begin{equation}
   P(z) = \int_0^{1} \int_0^1 db_i dd_i \, \delta \left((b_i-sd_i)-z \right).
 \end{equation}
 A careful treatment of the limits of integration yields,
 \begin{equation}
 \begin{split}
  P(z) =
  \left\{\begin{array}{ll}
   {(1-z)/s}, & 0 \le z \le 1,\\
          {1/ s},  &1-s \le z \le 0,\\
          1+ {z / s},  &-s \le z \le 1-s.
         \end{array}\right.
 \end{split}
 \end{equation}
  and $P(z)=0$ outside.
  This distribution has the shape of a trapezoid and $P(0)=1/s$.
 (It follows that $\left<z\right>=(1-s)/2$.)
 This implies that when $(b_j-sd_j) \to 0$, the probability that $(b_i-sd_i) \to 0$ for any $i \ne j$ is not suppressed. Furthermore, the individual  $(b_i-sd_i) \to 0$ approach are more or less independent, so the $n=N_{\rm cs}$th order zero is within the parameter ranges we are interested in and we may intuitively expect a peaking behavior 
 \begin{equation}
 \label{mc7peak}
  P(w_0) \propto \left( \ln |w_0| \right)^{N_{\rm cs}-1} \ {\rm as} \  \quad w_0 \to 0.
 \end{equation}

\item Let $r_i=d_i/b_i$. At the symmetry point where all $r_i$ are equal, $r_1=r_2=. . . =r_{N_{\rm cs}}=r$, we see that 
$W_0|_{\rm min} ({N_{\rm cs}})$ (\ref{mc7b}) collapses to
 \begin{equation}
   W_0|_{\rm min} ({N_{\rm cs}})= -  \frac{2(c_1+ s c_2)(1-sr)}{{N_{\rm cs}}(1+sr)}.
  \end{equation}
  As $r$ sweeps pass $1/s$, we obtain only a single zero in $w_0$. For this symmetry point, we expect $P(w_0)$ to be smooth at $w_0=0$. That is, the peaking behavior (\ref{mc7peak}) will disappear if we restrict ourselves to the symmetry point. So if  the peaking behavior in $P(w_0)$ is to survive, it must come from the vicinity of the symmetry point, which has a bigger region of parameter space than that of the symmetry point.  Away from this symmetry point, some zeros in the numerator of $w_0$ (\ref{mc7b}) may be canceled by coincident zeros in the denominator of $w_0$ (\ref{mc7b}), depending on the choice of the random parameters. This may lead to some transient behavior in  $P(w_0)$.      
Solution for $s$ at the symmetry point is discussed in appendix \ref{sec:discussion-solving-s}.
  
  \item   Constraint among the parameters comes into play as we limit the range of $s$ so the weak coupling approximation is valid. This will also introduce some transient behavior in $P(w_0)$. This transient behavior should fade away as we go to large $N_{\rm cs}$.

  \item Note that both the expression for $s$ (\ref{mc7}) and the expression for $w_0$ (\ref{mc7b}) are invariant under the simultaneous rescaling $b_i \to \lambda_ib_i$ and $d_i \to \lambda_id_i$, so $P(w_0)$ should be unchanged when we change the range of both $b_i$ and $d_i$ by a factor of $\lambda_i$. 
        Recall the example  (\ref{zexp1}) where the expectation value strongly depends on the range of the  random variables. 
        This would be very unsatisfactory here since we do not know the reasonable ranges for $(b_i, d_i)$.
        Fortunately, because of this $\lambda_i$ scaling property, we see that both $P(w_0)$ and $\left< w_0^p \right>$ are insensitive to the rescaling of the $(b_i, d_i)$ ranges.
        We shall find some reasonable choices of ranges for $b_i$ when we come to the potential $V$ and $P(\Lambda)$. Here, we restrict ourselves to $w_0<0$ so we simply need to look at only $\left<w_0\right>$ (instead of $\left<w_0^2\right>$). 
  Suppose we take the range of $c_i$ to be $[-{\rm r}_{\rm cutoff}, + {\rm r}_{\rm cutoff}]$, then $\left<w_0\right>/{\rm r}_{\rm cutoff}$ is independent of ${\rm r}_{\rm cutoff}$ and the ranges of $b_i, d_i$. For example, for $N_{cs}=5$, we have  $\left<w_0\right>/{\rm r}_{\rm cutoff} \simeq -2$.

\end{itemize}

Overall, we expect that $P(w_0)$ peaks more sharply at $w_0=0$ as $N_{\rm cs}$ increases; however, we do not know analytically the precise peaking behavior of $P(w_0)$.

So we expect that the expectation value  $\left< w_0 \right>$ decreases as $N_{\rm cs}$ increases. Because of the nontrivial forms of $s$ (\ref{mc7}) and $w_0$ (\ref{mc7b}), it is crucial to find their behaviors numerically.

  \subsection{Numerical comparison of $w_0$}

  In the above analysis, we see the product type of structure in $W_0|_{\rm min}$ (\ref{mc7b}). 
  However, since the parameters in the denominator are not independent of those in the numerator,  it is important to 
 perform numerical analysis and check whether the distribution does become more sharply peaked as we increase the number of complex structure moduli.

  Now we impose a randomness in the model (\ref{mc1}).
  Given random values for $c_1, c_2, b_i, d_i$ obeying uniform distribution, we solve supersymmetric conditions $D_S W_0 = 0, D_{U_i} W_0 = 0$ for real part of moduli fields, $s, u_i$.
  Here we restrict ourselves in $-1 \leq c_1, c_2, b_i, d_i \leq 1$ for just a choice.
  We are also interested in the region where string coupling $g_s = e^\phi = s^{-1}$ can be treated perturbatively, therefore we further impose $s>1$ together with $u_i>0$
  (which  is required for a real-valued K\"ahler potential in our definition).
  
  Recall that the $w_0 A_1 <0$ condition is required for the K\"ahler moduli stabilization with positive $\Lambda$ as we see in section \ref{sec:revi-analyt-study}. Because of the reflection symmetry, we may focus on the $w_0<0$ region without losing generality here.
  It is worth commenting that all of the numerical solutions obtained just by solving the supersymmetric condition actually satisfy the positivity of the Hessian. Thus we can insert the solution in the model for the K\"ahler moduli stabilization.

  Inserting the solutions for $s$ (\ref{mc7}) and $u_i$ (\ref{mc4}) into $w_0=W_0|_{\rm min}$ (\ref{mc7b}), we obtain the distribution $P(w_0)$  and its behavior.
  In Figure \ref{figw_0}, we show the resultant probability distribution $P(w_0)$ for various $h^{2,1}=N_{\rm cs}$, where the dilaton dependence has been taken into account.
  Because of the non-trivial dependence in the denominator of (\ref{mc7b}), the peak at $w_0=0$ is smoothed out around $N_{\rm cs} = 3,4$, but emerges at $N_{\rm cs} \ge 5$.
  For $N_{\rm cs} \ge 5$, the peak at $w_0=0$ becomes sharper as we increase $N_{\rm cs}$.
  Although $W_0$ (\ref{mc1}) takes the form of a sum of terms and therefore one might expect a sum distribution (which is smooth at $W_0=0$, as shown in Figure \ref{randomW0}), instead of a product distribution (which typicaly peaks at $0$), the data clearly shows a sharper peak in $P(w_0)$ (Figure \ref{figw_0}) as we increase the number of complex structure moduli. This is nothing but the outcome of the correlation of the terms through the stabilization of complex structure moduli and dilaton.

  \begin{figure}[t]
   \begin{center}
    \includegraphics[width=13em]{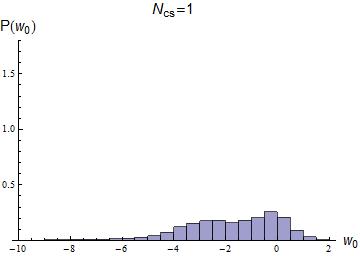}
    \includegraphics[width=13em]{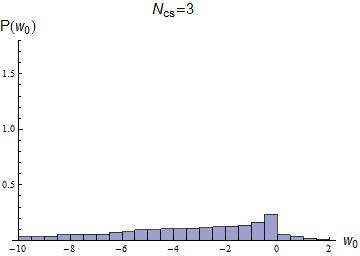}
    \includegraphics[width=13em]{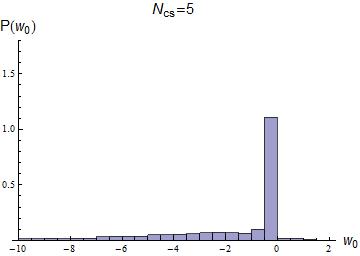}
    \includegraphics[width=13em]{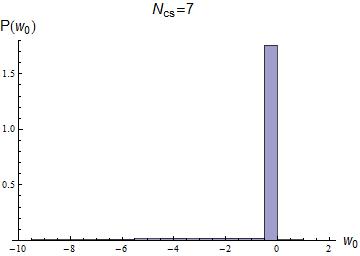}
    \includegraphics[width=15em]{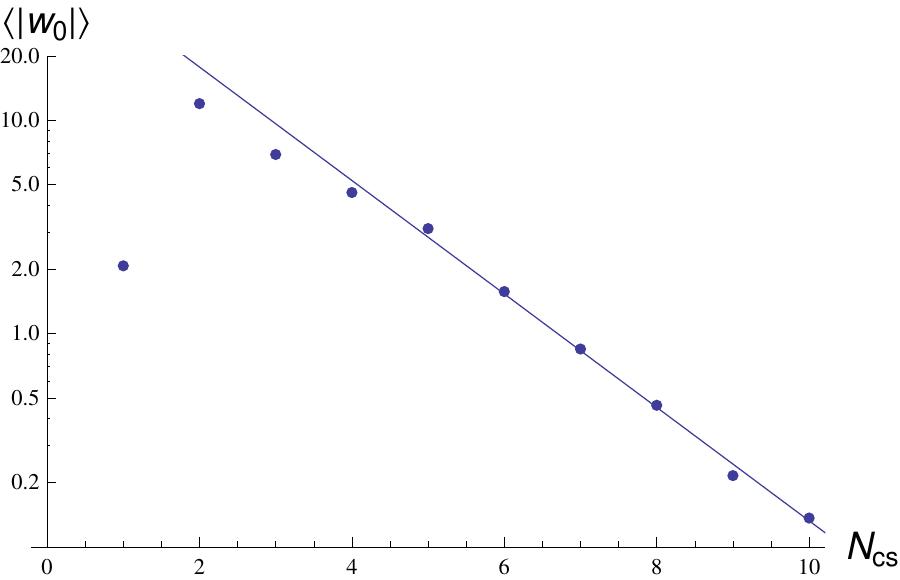}
   \end{center}
   \caption{\footnotesize The probability distribution $P(w_0)$ for $-10 \le w_0 \le 2$ with fixed $b_i=-1$ at $N_{\rm cs}=1, 3, 5, 7$ respectively.
   In the last figure, we show the behavior of the  expectation value $\left< |w_0 |\right>$ for $N_{\rm cs} \le 10$.
   }
   \label{fig:W0-fixb-comparison}
  \end{figure}

Although the distribution feature for the parameters $c_1, c_2, b_i, d_i$ are reasonably well-motivated from the flux compactification because these parameters are from the 3-form fluxes $H_3, F_3$, we like to consider other possibilities.
Since $u_i= w_0/2b_i(1-sr_i)$, allowing $b_i=0$ implies that we are looking at vacua where some $u_i \to \infty$, which may not be valid within our approximation. So, as an alternative, let us restrict the values of $b_i$ to a range that excludes zero, say $0 > -f_{min} \ge b_i \ge -f_{max}$. It turns out that the qualitative physics does not change if we simply restrict $b_i$ to a fixed value, say $b_i= -f<0$, where we may consider $-f$ to be the average value of $b_i$ with a range.
Note that $u_i$ stays finite as $(1-sr_i) \to 0$, since $w_0 \to 0$ in this limit as well.

Again, we can solve for $s>1$ and $u_i>0$ and determine $w_0$.
In Figure \ref{fig:W0-fixb-comparison}, we show the distributions of $P(w_0)$ with fixed $b_i=-1$ while allowing the other parameters uniformly distributed with the range $-1\leq c_1 ,c_2, d_i \leq 1$.
We again impose the weak string coupling constraint $s>1$ as well as $u_i>0$.
As apparent from the comparison between Figure \ref{figw_0} and Figure \ref{fig:W0-fixb-comparison}, the distribution of $w_0$ in this case is more peaked toward $w_0 =0$.
Note that the distribution for $w_0 >0$ is quite different from that for $w_0\leq 0$ in this setup.
 This is because the sign fixing of $b_i$ violates the reflection symmetry of $w_0$ (realized by flipping signs of all the parameters $c_i, b_i, d_i$ simultaneously in the model).

We also estimate the expectation value $ \left< |w_0| \right> $ as a function of $N_{\rm cs}$, especially for $N_{\rm cs}= 4-10$. In an attempt to extract the large  $N_{\rm cs}$ behavior, we neglect the data points for $N_{\rm cs} < 4$, because of the transient behavior at lower $N_{\rm cs}$. The data points (for $N_{\rm cs} \geq  4$) in Figure \ref{fig:W0-fixb-comparison} are well fitted by
$  \left< |w_0| \right> = 60.6 \, e^{-0.613 N_{\rm cs}} $.
However, numerical data indicates that the drop of  $\left< |w_0| \right>$ as a function of $N_{\rm cs}$ slows down as we go to $N_{\rm cs} >10$ (see LHS of Figure \ref{values-for-peaking}).

The peaking behavior of $P(|w_0|)$ may be quantified via a comparison of the values at a fixed percentage of $|w_0|$ accumulating from the zero point. Let us introduce $|w_0|^{Y\%}$ defined by $\int_0^{|w_0|^{Y\%}} P(|w_0|) \, d \, |w_0| = Y\%$. That is, there is a $Y \%$ chance that $|w_0|$ will be smaller than  $|w_0|^{Y\%}$.
In LHS of Figure \ref{values-for-peaking}, we show $\left< |w_0| \right>$, $|w_0|^{80\%}$ and $|w_0|^{10\%}$ as a function of 
$N_{\rm cs} $ in the range $1\leq N_{\rm cs} \leq 30$. We see that, for $N_{\rm cs} > 3$, they all decrease as the number of complex structure moduli increases.

  \subsection{Inserting the $w_0$ solution into the single K\"ahler modulus model \label{sec:plugg-single-kahl}}

  Now we consider the moduli stabilization in the presence of single K\"ahler modulus, given the data of $w_0, s, u_i$ obtained by solving supersymmetric conditions for complex structure and dilaton moduli.
  The potential we have in mind here is given by
  \begin{equation}
   \begin{split}
    K =& -2   \ln \left({\cal V} + {\hat{\xi} \over 2} \right) - \ln (2s) - \sum _i^{h^{2,1}} \ln (2u_i), \\ 
    {\cal V} =& \gamma_1 (T_1 + \bar{T_1})^{3/2}, \quad 
    \hat{\xi} = - {2 \zeta (3) \over 3 (2\pi)^3} \, s^{3/2} \, 2(h^{1,1}-h^{2,1}),  \\
    W =& w_0  + A_1 e^{-a_1 T_1}.
   \end{split}
   \label{potential for the combination}
  \end{equation}
  We consider the stabilization of single K\"ahler moduli with the potential above, under the assumption of complex structure moduli stabilization at higher energy, which is not affected by the next leading order corrections for K\"ahler modulus stabilization. Again, we choose the solution where ${\rm Im }  T_1=0$ and focus on $t_1 = {\rm Re }  T_1$.
  The potential above has additional contribution in the overall factor in addition to (\ref{eq:potential for single Kahler}), which is estimated by
  \begin{equation}
   \begin{split}
    {V  } \sim& {-w_0 a_1^3 A_1 \over 2 \gamma_1^2 2^{N_{\rm cs}+1} s \prod u_i} \left({2C \over 9 x_1^{9/2}} - {e^{-x_1} \over x_1^2} \right),\\
    C = & {-27 w_0 \hat{\xi} a_1^{3/2} \over 64 \sqrt{2} \gamma_1 A_1}, \quad
    x_1 = a_1 t_1,
   \end{split}
   \label{approx potential with complex}
  \end{equation}
  where we approximate the potential up to linear order of ${\hat{\xi}/ {\cal V}}, A_1e^{-x_1}/ W_0$, and $N_{\rm cs} = h^{2,1}$ just for convenience.
  Here, we set $a_1=\gamma_1=1$ for simplicity.

  In this section, we present the data obtained for the following 4 cases :
  \begin{enumerate}
   \item $-1 \leq c_1,\, c_2 \leq 1, \quad -f_1 \leq b_i,\, d_i \leq f_1,$
   \item $-1 \leq c_1,\, c_2,\, r_i={b_i / d_i} \leq 1 \quad  -f_2 \leq b_i \leq f_2,$
   \item $-1 \leq c_1,\, c_2 \leq 1,  \quad b_i=\pm f_3, \quad -f_3 \leq d_i \leq f_3, $
   \item $-1 \leq c_1,\, c_2 \leq 1,  \quad b_i=-f_4, \quad -f_4 \leq d_i \leq f_4, $
  \end{enumerate}
  with uniform distributions for the ranges of the random parameters as shown. Note that we impose uniform distribution for $r_i$ instead of $d_i$ in Case 2.
  The positively defined parameter $f_j$ will be fixed later.
  For each case, we insert the given distributions into (\ref{simplest model for complex}) and solve for  the supersymmetric conditions (\ref{SUSY1}) for $s, u_i, w_0$, and collect sets of data for $s>1,\,  u_i>0$.
  The invariance under rescaling $b_i, d_i$ simultaneously in (\ref{mc7b}), the peaking behavior of $w_0 \leq 0$ as well as the distribution itself is independent of $f_j$.

  As we have discussed in the previous subsection, we observe the peaking behavior in $w_0$ as $N_{\rm cs}$ increases.
  However, we also see that $P(u_i)$ are also peaked toward $u_i=0$, and therefore the coefficient of the potential $V$(\ref{approx potential with complex}) goes essentially infinity, meaning that the resultant $\Lambda$ itself may diverge.
  Since we use the effective potential for the 4D gravity analysis, the maximal value of the potential should be within the Planck scale for it to be valid within the approximation used.
 Since $V(x_1=0)$ blows up, it makes more sense to use the barrier height as a measure of the scale of the potential.
 The easiest way to implement this constraint is for the barrier height (\ref{barrier}) to be no higher than the Planck scale, or
  \begin{equation}
   V_H \equiv  {1\over 9} \left({2 \over 5}\right)^{9/2}(0.24){-w_0 A_1 \over 2^{N_{\rm cs}+1} s \prod u_i} \leq 1.
   \label{V_H}
  \end{equation}
  That is, we keep only those $V_H$ that are less than unity. Here we give $A_1$ an uniform distribution in the range $-1 \leq A_1 \leq 1$.
 To enhance the data sample without changing the underlying physics, we may simply restrict ourselves to small values of $f_i$, where $V_H \propto f_j^{N_{\rm cs}}$.
There may be a number of different ways to choose the cutoff point. 
  Here we employ a specific way to achieve this: we count the number of data starting from $V_H=0$, and we choose $f_i$ so that $90\%$ of $V_H \leq 1$, then we set this as the cutoff point and keep only this $V_H \leq 1$. This implies that we restrict ourselves to the cases where $\Lambda \lesssim 1$.
  Note that we do not lose generality by the restriction $w_0 \leq 0$ for Case 1-3, since we have the reflection symmetry of $w_0$ by changing the signs of all parameters $c_i, b_i, d_i$, which remains solutions for $s, u_i$ unchanged.
Changing the $90\%$ cutoff to a higher or slightly lower value does not change the qualitative picture.

  \begin{figure}
   \begin{center}
    \includegraphics[width=17em]{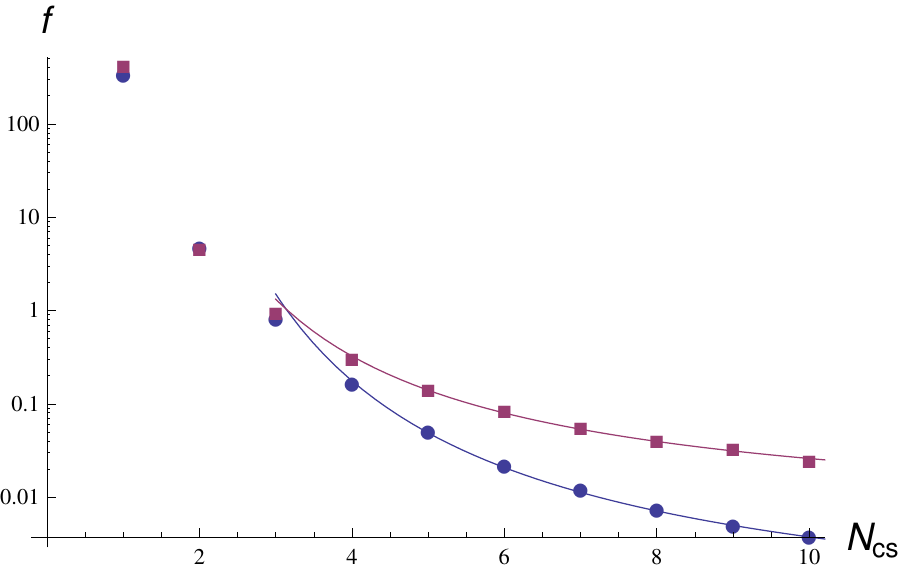}
    \includegraphics[width=17em]{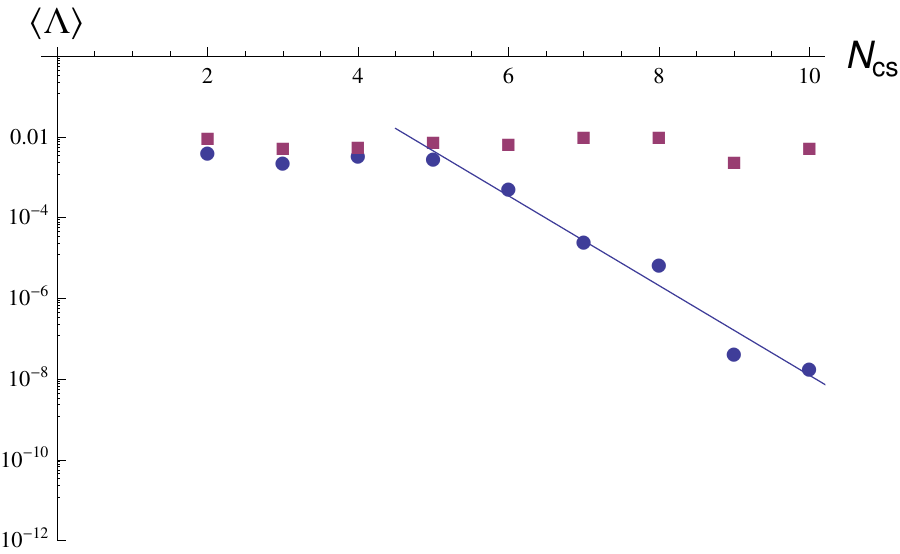}
   \end{center}
   \caption{\footnotesize The estimated values of $f_k$ and $\left< \Lambda \right>$ in Case 1 (purple square) and Case 4 (blue circle).
   The fitting curves here are $f_1 \sim e^{-5.33 + 16.9/N_{\rm cs}}$ (red on LHS), $\ f_4 \sim e^{-8.16 + 25.7/N_{\rm cs}}$ (blue on LHS) and $\left< \Lambda \right>_4 \sim e^{-2.56 N_{\rm cs} + 7.40}$ (blue on RHS). However, the drop of $\left< \Lambda \right>$ as a function of $N_{\rm cs}$ slows down appreciably for larger values of $N_{\rm cs}$.}
   \label{fig:case1}
  \end{figure}

  Let us estimate $f_j$ in each case. To keep $90\%$ of $V_H \le 1$, we need to choose a smaller $f_j$ as $N_{\rm cs}$ increases.
In LHS of Figure \ref{fig:case1}, we show the plot for the value of $f_j$ required to normalize the cutoff point of $V_H$ in Case 1, 4 just for illustration.
  For $N_{\rm cs}\geq 4$, $f_j$ may be estimated to be :
  \begin{equation}
   f_1 \sim e^{-5.33 + 16.9/N_{\rm cs}}, \quad
    f_2 \sim e^{-4.62 + 20.5/ N_{\rm cs}}, \quad
    f_3 \sim e^{-5.78 + 19.6/ N_{\rm cs}}, \quad
    f_4 \sim e^{-8.16 + 25.7/ N_{\rm cs}}.
  \end{equation}
We see that $f_i \ll 1$ at larger $N_{\rm cs}$.

  Now we are ready to plug the data into (\ref{approx potential with complex}).
  Since we consider just the single K\"ahler modulus stabilization with the sets of data inputs, the stabilization mechanism follows that reviewed in section \ref{sec:revi-analyt-study}, where the combined parameters obey $3.65 \lesssim C < 3.89$ for metastable $dS$ vacua.
  First we give $A_1$ an uniform distribution with range  $-1 \leq A_1 \leq 1$. Next we choose the data for one of the four cases mentioned above for each fixed $N_{\rm cs}$. 
  The probability distribution $P(\Lambda)$ of $\Lambda$ in Case 1 is illustrated in Figure \ref{figLambda}.
  We see that there is the clear peaking behavior toward $\Lambda =0$, which is getting sharper as $N_{cs}$ increases. 
  However, we do not see any clear trend for $\left< \Lambda \right>$ in RHS of Figure \ref{fig:case1}.
  This is probably because there is a non-trivial long tail in large $\Lambda$ in $P(\Lambda)$ which contribute significantly to 
$\left< \Lambda \right>$.
In Case 2 and 3, we see a clear peaking of $P(\Lambda)$ at $\Lambda=0$, similar to that in Case 1; but again $\left< \Lambda \right>$ behaves quite similarly to that in Case 1, without a clear trend one way or another.

  Next, let us consider Case 4, with $b_i=-f_4$ and $w_0 \lessgtr 0$.  The result is shown in Figure \ref{fig:case1}. The peaking behavior of $P(\Lambda)$ at $\Lambda=0$ is again evident, while $\left< \Lambda \right>$ has a rather different behavior than the other 3 cases.  The $\left< \Lambda \right>$ in Case 4 is shown in Figure \ref{values-for-peaking}. We find that it is roughly given by
  \begin{equation}
   \left< \Lambda \right>_4 \sim e^{- 2.56 N_{\rm cs} + 7.40}
  \end{equation}
  for $N_{\rm cs} \le 10$. However, the drop in $ \left< \Lambda \right>$ slows appreciably for $N_{\rm cs} > 10$ as showed in RHS of Figure \ref{values-for-peaking}. 
Still, this example illustrates the sensitivity of $\left< \Lambda \right>$ to the input ranges and distributions of the random parameters.

Similarly to $w_0$, we can quantify the peaking behavior of $P(\Lambda)$ by estimating the value at a fixed percentage of data counting from $\Lambda = 0$. 
Let us introduce $\Lambda^{Y\%}$, defined by
$\int_0^{\Lambda^{Y\%}} \, P(\Lambda) \, d\Lambda = Y\%$. That is, there is  a $Y\%$ chance that $\Lambda$ will fall in the range $\Lambda^{Y\%} \ge \Lambda \ge 0$.  In Figure \ref{values-for-peaking},  we show $\left< \Lambda \right>$, $\Lambda^{80\%}$ and $\Lambda^{10\%}$ as a function of the number of complex structure moduli. For example, at $N_{\rm cs}(\equiv h^{2,1})=20$, while $\left< \Lambda \right> \simeq 1.2 \times 10^{-8}$, we have
$\Lambda^{80\%} \simeq 2.0 \times 10^{-19}$ and $\Lambda^{10\%} \simeq 4.8 \times 10^{-28}$. 
Although the drop of $\left< \Lambda \right>$ is quite mild after $N_{\rm cs} > 10$, we see that $\Lambda^{80\%}$ and  $\Lambda^{10\%}$ continue to drop appreciably.
This means that most $\Lambda$ have small values, while there exist a long tail $\Lambda \lesssim 1$ in the distribution $P(\Lambda)$. 
At $N_{\rm cs}=30$,  $\Lambda^{80\%} \simeq  3.61 \times 10^{-28}$. Extrapolating to higher $N_{\rm cs}$, we find that the likely value of $\Lambda$ (here we mean $\Lambda^{80\%}$) will be small enough for  $N_{\rm cs} \sim 140$, which is in the ballpark of manifolds studied.

\begin{figure}[t]
 \begin{center}
  \includegraphics[width=15em]{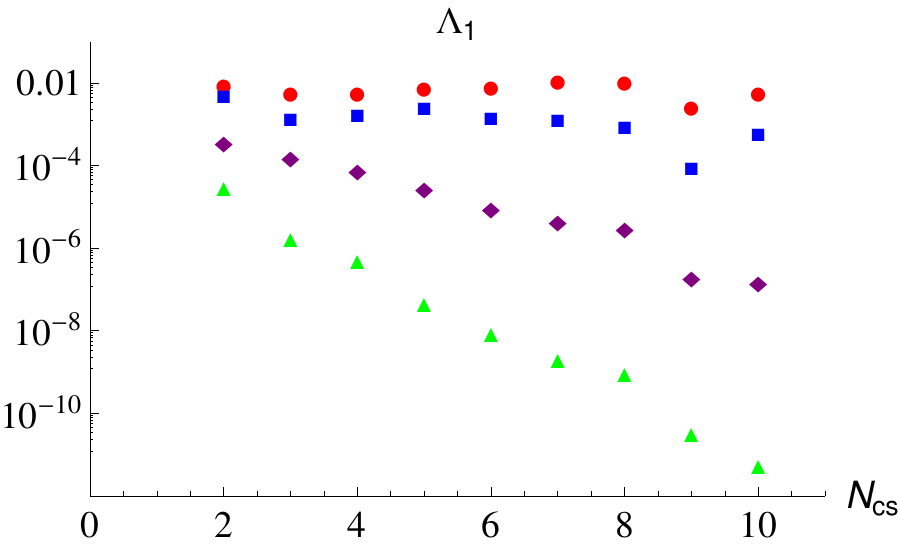}
 \end{center}
 \caption{\footnotesize The comparison of $\left<\Lambda\right>_1$ (red circle),  $\Lambda_1^{80\%}$ (blue square), $\Lambda_1^{50\%}$ (purple diamond), and $\Lambda_1^{10\%}$ (green triangle) between $N_{\rm cs} = 2-10$ for the data of Case 1 with most general random inputs. $\left<\Lambda\right>_1$ is the same data set as that (purple squares) shown in Figure \ref{fig:case1}.} 
 \label{fig:Lambda-values-randomb}
\end{figure}

Let us also find out $\Lambda_1^{Y\%}$ for the data of Case 1 shown in Figure \ref{fig:case1}.
Here we just show data up to $N_{\rm cs} =10$ since computational cost is huge for $N_{\rm cs} >10$.
In Figure \ref{fig:Lambda-values-randomb}, we see clear decreasing behaviors in $\Lambda_1^{10\%}$ and 
$\Lambda_1^{50\%}$, but not in $\left<\Lambda\right>_1$ or $\Lambda_1^{80\%}$. 
 
By choosing smaller $|b_i|$ allows us to accumulate more data.
What happens if we impose a different cutoff on $V_H$.
  As an example, let us choose $b_i$ so that, instead of $90\%$, only $20\%$ of the potentials have $V_H \le1$. 
That is, the potentials are less squeezed to small values in the $20\%$ cutoff scenario.
In this case, we find that, for $N_{\rm cs}=20$, $\Lambda^{80\%} \simeq 8.6 \times 10^{-6}$ and $\Lambda^{10\%} \simeq 1.2 \times 10^{-9}$, which are substantially bigger than those for the $90\%$ cutoff case. Clearly we have to understand this squeezing issue better.

Although we have employed the cutoff such that $90\%$ of the data is within the Planck scale (as measured by the barrier height $V_H$) to deal with its divergent behavior at each $h^{2,1}$, there are other ways to impose the cutoff.
For instance, we may consider a fixed cutoff $f_1 = 0.1$ in Case 1 at all $h^{2,1}$, where we still have enough statistics within Planck scale for $h^{2,1} \leq 10$.
Then the likely values $\Lambda^{Y\%}$ are actually not dropping exponentially.
So we see that the physics depends on the details of the way in implementing a cutoff, though the previous $90\%$ cutoff is motivated by the statistical analysis of data.

So far we have considered the physical quantities $\Lambda$ and the related potential barrier $V_H$ in this subsection.
However, one may also consider, instead of $V_H$,  the quantity $e^{K_{\rm cs+d}} |W_0|^2$, which is invariant under the K\"ahler transformation.
The K\"ahler invariant quantity at a minimum is given by $e^{K_{\rm cs+d}} |W_0|^2 = w_0^2/(2^{N_{\rm cs}+1} s \prod u_i)$.
Since this quantity may diverge due to the $u_i$ in the denominator (as in the case of $V_H$), we need to introduce the cutoff to make this quantity within Planck scale (this quantity is a term in the potential).
If we follow the same procedure as that for $V_H$, the distribution of the K\"ahler invariant quantity at each $N_{\rm cs}$ suggests sharply peaked distribution, which is qualitatively similar to what we have seen already for $P(\Lambda)$.

\subsection{Some remarks}

A few remarks may be in order here: 

\begin{itemize}

 \item Note that the ``product'' form emerges in the  superpotential value $w_0$ (\ref{mc7b}), despite the fact that the parameters $b_i$ and $d_i$ appear in the ``sum'' form in the original superpotential (\ref{simplest model for complex}). As pointed out in \cite{Sumitomo:2012wa}, any linear combination of a set of random variables, as naively the case in  (\ref{simplest model for complex}), will not yield a peaking behavior.
That is, $W_0$ will be smooth at zero if the $u_i$, the real part of $U_i$ in $W_0$ are treated as independent arbitrary constants.
      When we go to the supersymmetric point for the minimum, $u_i$  become functions of $b_i$ and $d_i$, which leads to the product form in (\ref{mc7b}). It is the (complex structure and dilation) moduli stabilization dynamics that converts the ``sum'' form (\ref{simplest model for complex}) to the ``product'' form (\ref{mc7b}), which is the key to the peaking property of $P(w_0)$ and $P(\Lambda)$, the latter of which is necessary for the stringy mechanism for a naturally small $\Lambda$ to work.

 \item The actual values of $u_i$ and $s$ will surely be shifted away from the supersymmetric point after supersymmetry breaking.
The correction of back reaction to the set of solutions is suppressed due to the suppression of large volume and larger $W_0$ compared to the non-perturbative terms if we work in this parameter region \cite{Rummel:2011cd}.
       Even if we include the corrections in $W_0$, the modification would be negligible and therefore the feature of peaked behaviors should remain intact.

 \item It is possible (even likely) that the coefficient $A_k$ in (\ref{LVS effective potential}) emerges as a consequence of the stabilization of higher scale moduli. In this case, we expect the probability distribution $P(A_k)$ to be peaked at $A_k=0$. This will enhance the peakiness of $P(\Lambda)$, so fewer complex structure moduli may be needed for a small enough $\Lambda$.

 \item  Both the model and its solution we study here are non-trivial but simple enough for a clear analysis. This opens some questions on the simplifications we are taking.
There are additional interaction terms in both $W$ and $K$ coming from higher order corrections.
       Based on the limited experience we have developed so far, one may hope that those (higher $\alpha'$ or stringy loop \cite{Cicoli:2007xp,Cicoli:2008va,Berg:2007wt}) corrections will lead to tighter interactions among the moduli and so will actually strengthen the peaking behaviors. However, this does not necessarily imply an exponentially small $\Lambda$.
    It is obviously interesting to study other stringy models.

\end{itemize}

\section{Multi-K\"ahler moduli cases \label{sec:multi-khaler-moduli} \label{sec:numer-analys-distr}}

A model for multiple K\"ahler moduli stabilization at positive $\Lambda$ is even suggested especially with non-perturbative effects in superpotential and $\alpha'$ correction in K\"ahler, considered in \cite{Balasubramanian:2004uy,Westphal:2006tn,Rummel:2011cd,deAlwis:2011dp} or known as {\it K\"ahler uplifting model}.
The model basically assumes the stabilization of the complex structure moduli and dilaton-axion moduli at high scale, while the K\"ahler moduli stabilization is achieved at low energy scale with the non-perturbative effect and the $\alpha'$-correction.
Together with the non-trivial constant $W_0$ in the superpotential, given after complex and dilaton moduli stabilization, we can have a small positive minimum as a result of K\"ahler moduli stabilization.

We consider the supergravity effective potential which is obtained after the Calabi-Yau compactification in type IIB:
\begin{equation}
 \begin{split}
  K =& -2   \ln \left({\cal V} +{\hat{\xi} \over 2} \right), \quad
  {\cal V} = \gamma_1 (T_1 + \bar{T_1})^{3/2} - \sum_{i=2}^{N_K} \gamma_i (T_i + \bar{T}_i)^{3/2},\\
  W =& W_0 + \sum_{i=1}^{N_K} A_i e^{-a_i T_i},
 \end{split}
 \label{effective potential for multi Kahler}
\end{equation}
where we have again introduced the non-perturbative terms proportional to $A_i$ for all of $N_K=h^{1,1}$ K\"ahler moduli, and the $\alpha'$ correction defined $\hat{\xi} =  -\zeta(3)\chi(M)/(4\sqrt{2}(2\pi)^3)$.
We basically require $\hat{\xi} >0$ for the stability analyses later, meaning that the number of complex structure moduli is greater than the number of K\"ahler moduli: $\chi(M)=2(h^{1,1}-h^{2,1})<0$.
$W_0$ comes from the flux contributions, and generically does not depend on K\"ahler moduli (no-scale structure).

We assume that the non-perturbative effect and the $\alpha'$ correction are small enough compared to the other quantities as well as in section \ref{sec:revi-analyt-study}, so that we analyze the potential up to linear orders of
\begin{equation}
 {{\hat{\xi}} \over {\cal V}}\ll 1,\quad \left|{A_i e^{-a_i T_i} \over W_0}\right|\ll 1.
  \label{assumptions in the potential}
\end{equation}
These assumptions are quite natural since the $\alpha'$ correction can be treated to be small in perturbation theory.
As we see in the previous section, this perturbative analysis agrees with the analysis for small $\Lambda$.
There are essentially two important outcomes of these assumptions: (1) suppression of the off-diagonal components between the K\"ahler components and the complex structure/dilaton components in the mass matrix, and (2) simplification of the potential for the K\"ahler moduli since we can neglect the next-to-next leading order in the potential with respect to the approximation (\ref{assumptions in the potential}).

We focus on the real part of K\"ahler moduli $\re T_i = t_i$ because the imaginary part always has a potential of cosine type and therefore  $\im T_i = 0$ always satisfy the extremal conditions.
The $\im T_i = 0$ solutions also decouple the real sector from the imaginary sector completely in the  mass matrix \cite{Rummel:2011cd}. Thus we can safely neglect the imaginary components.

The potential is expanded with respect to (\ref{assumptions in the potential}), so
\begin{equation}
 \begin{split}
  {V  } \sim& {3 W_0^2 \hat{\xi} \over 64\sqrt{2} {\cal V}^3} + \sum_{i=1} {a_i t_i A_i e^{-a_i t_i} W_0 \over 2 {\cal V}^2}.
 \end{split}
\end{equation}
Defining the following parameters:
\begin{equation}
 \begin{split}
  &x_i = a_i t_i,  \quad 
  \delta_i = {\gamma_i a_i^{-3/2} \over \gamma_1 a_1^{-3/2}}, \quad 
  C = {-27 W_0 \hat{\xi} a_1^{3/2} \over 64 \sqrt{2} \gamma_1 A_1}, \quad
  B_i ={A_i \over A_1},
 \end{split}
 \label{eq:combined parameters}
\end{equation}
the potential takes the approximate form:
\begin{equation}
 \begin{split}
  {V } \sim& - {A_1 \, W_0  a_1^3 \over 2 \gamma_1^2 }
  \left( {2 C \over 9 (x_1^{3/2} - \sum_{j=2} \delta_j x_j^{3/2} )^3} - {x_1 e^{-x_1} \over (x_1^{3/2} - \sum_{j=2} \delta_j x_j^{3/2} )^2} \right. \\
  & \hspace{18em} \left. - \sum_{i=2} {B_i x_i e^{-x_i}\over (x_1^{3/2} - \sum_{j=2} \delta_j x_j^{3/2} )^2} \right).
 \end{split}
 \label{rewritten potential for Kahler}
\end{equation}
Here, we have a symmetry of exchanging moduli fields $x_i$ together with the assigned parameters $B_i$, coming from the symmetry for $T_i\, (i\neq 1)$ which the full potential (\ref{effective potential for multi Kahler}) contains.

 Our purpose here is twofold :  to study (1) the effect of increasing the number $h^{1,1}$ of K\"ahler moduli and (2)  the effect of different distributions for $W_0$ and $A_i$, on both $P(\Lambda)$ and $\left<  \Lambda \right> $. For the first case, we shall simply treat  $W_0, A_i$ as random parameters with uniform distributions.

  \subsection{Uniformly distributed $W_0, A_i$}

  Now we find the meta-stable minimum of $V$ (\ref{rewritten potential for Kahler}) to obtain $\Lambda$ and then estimate its probability distribution $P(\Lambda)$ and  $\left<  \Lambda \right> $.
  Here again we set $a_i = \gamma_i = \hat{\xi} =1$ to see just the effect of randomness of $W_0, A_i$ on $P(\Lambda)$
 and  $\left<  \Lambda \right> $. We shall start with uniform distributions for the parameters $W_0, A_i$.
Here we compare several different distributions with different setups to see how $P(\Lambda)$
 and  $\left<  \Lambda \right> $ behave.

  First we assume uniformly distributed random parameters $W_0, A_i$ with range: $-15 \leq W_0 \leq 0, \quad 0\leq A_i \leq 1$ for convenience.
  The choice of sign here is for increasing chance of $dS$ solutions.
  Even if we use the different choices, the resultant behavior is unchanged.
  The calculation cost for solving for $t_i$ and checking the stability can be hugely reduced if we limit ourselves in the parameter region which is valid for meta-stable vacua at $\Lambda \ge 0$.
  We follow the analysis of the parameter region up to three-moduli in appendix \ref{sec:stab-two-three}, and further apply the similar limitation for the models with more K\"ahler moduli.
  The limitation of our interest is the one we obtain in the case with three K\"ahler moduli, but we can expect the similar validity region for $B_i - B_j$ plane ($i \neq j$) as that of $B_2 - B_3$ plane, owing to the symmetry.
  This limitation hugely reduces our calculation costs so we can numerically get the statistical distributions of the potential (\ref{effective potential for multi Kahler}) even at larger numbers of K\"ahler moduli.

  \begin{figure}[t]
   \begin{center}
    \includegraphics[width=16em]{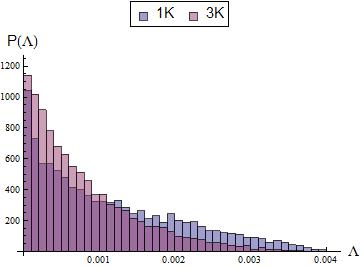}
    \includegraphics[width=20em]{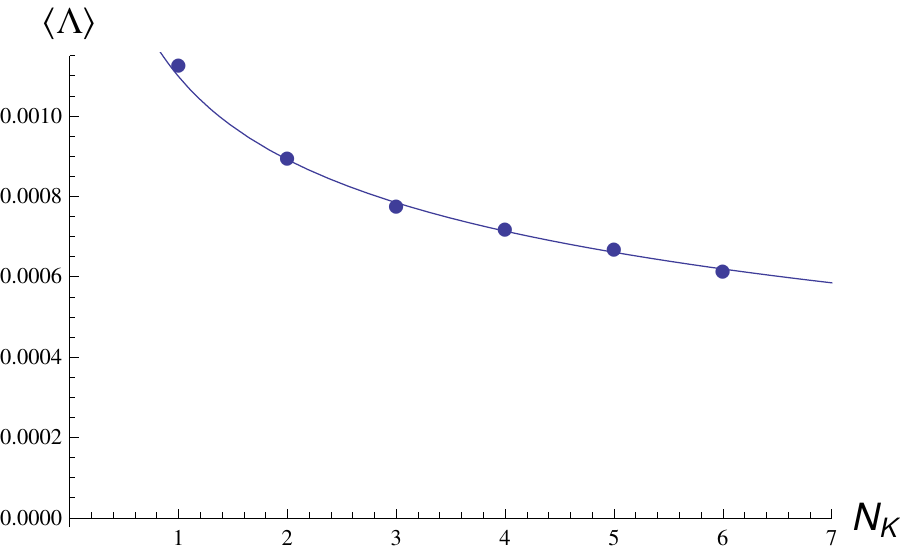}
   \end{center}
   \caption{\footnotesize The probability distribution $P(\Lambda)$ of $ {\Lambda}$ assuming uniformly distributed $W_0, A_i$.
    $\left<  {\Lambda} \right>$ as a function of the number $N_K$ of K\"ahler moduli; the curve is given by $\left<  {\Lambda}\right> \sim 0.00110 N_K^{-0.226} e^{-0.0268 N_K}$.}
   \label{fig:ccdist-uniform}
  \end{figure}

  In Figure \ref{fig:ccdist-uniform}, we show plots of the statistical analyses.
  If we assume uniformly distributed $W_0, A_i$, then the probability distribution $P(\Lambda)$  of $ {\Lambda}$ become more sharply peaked as the number $N_K$ of K\"ahler moduli increases.
  The plot in RHS suggests that the expectation value $\left<  \Lambda \right> $ of ${\Lambda}$ is suppressed as a function of $N_K$ :
  \begin{equation}
  \left< {\Lambda}\right> \sim  0.00111 N_K^{-0.282} e^{-0.0138 N_K},
   \label{<Lambda> for multi-Kahler}
  \end{equation}
  where we neglected the point at $N_K =1$ in estimation, since the single modulus model does not have off-diagonal components in the Hessian, and therefore we expect that the value at $N_K=1$ is not necessary to obey the estimated curve from the data among $N_K = 2-6$.

  Although we see the suppression of $\left<  \Lambda \right> $ as increasing $N_K$, we may worry that the similar suppression happens even for mass terms, because the mass matrix is also given as a complicated function of the parameters and moduli.
  We now estimate the physical mass matrix, which is the mass matrix for canonically defined fields.
  The physical mass matrix is related to the second derivative of potential $\partial_{t_i} \partial_{t_j} V$ through the diagonalization matrix of K\"ahler metric.
  The diagonalization matrix $D_{ij}$ of the K\"ahler metric is defined to satisfy
  \begin{equation}
   D_{ki} K_{ij} D^{T}_{jl} = \lambda_k \delta_{kl}
  \end{equation}
  where $\lambda_i$ are the eigenvalues of K\"ahler metric and same indeces in LHS are summed over.
  Then the canonical coordinate $Y_i$ is defined to be
  \begin{equation}
   d Y_i = \sqrt{2 \lambda_i} (D^{T})^{-1}_{ik} d t_k.
  \end{equation}
  Now we know the conversion matrix to physical mass matrix from the second derivative matrix:
  \begin{equation}
   \begin{split}
     {m^2_{ij}  } = {dt_k \over dY_i} {dt_l \over dY_j} \left.{\partial_{t_k} \partial_{t_l} V  }\right|_{\rm min}.
   \end{split}
  \end{equation}
  Performing the conversion at each stable point, we can calculate the physical mass matrix accordingly.

 \begin{figure}[t]
   \begin{center}
    \includegraphics[width=20em]{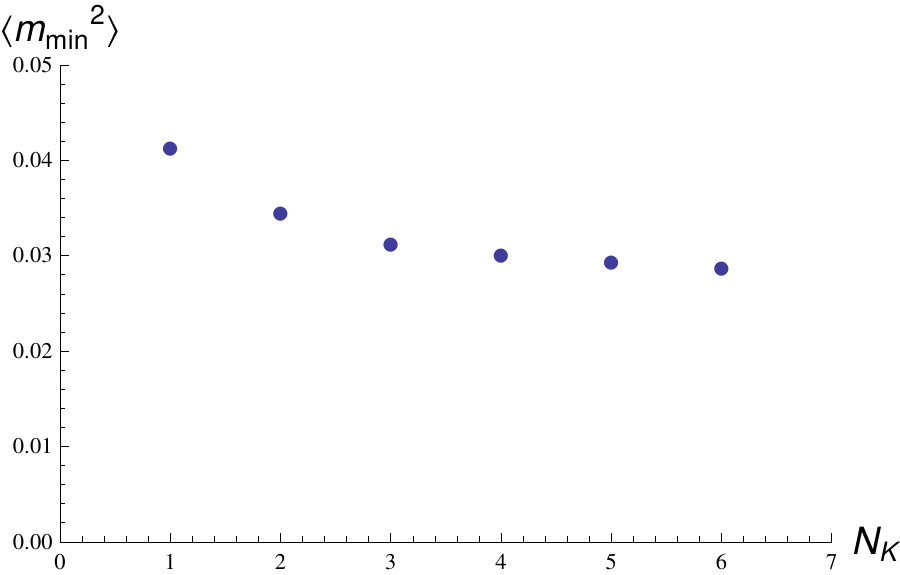}
    \includegraphics[width=20em]{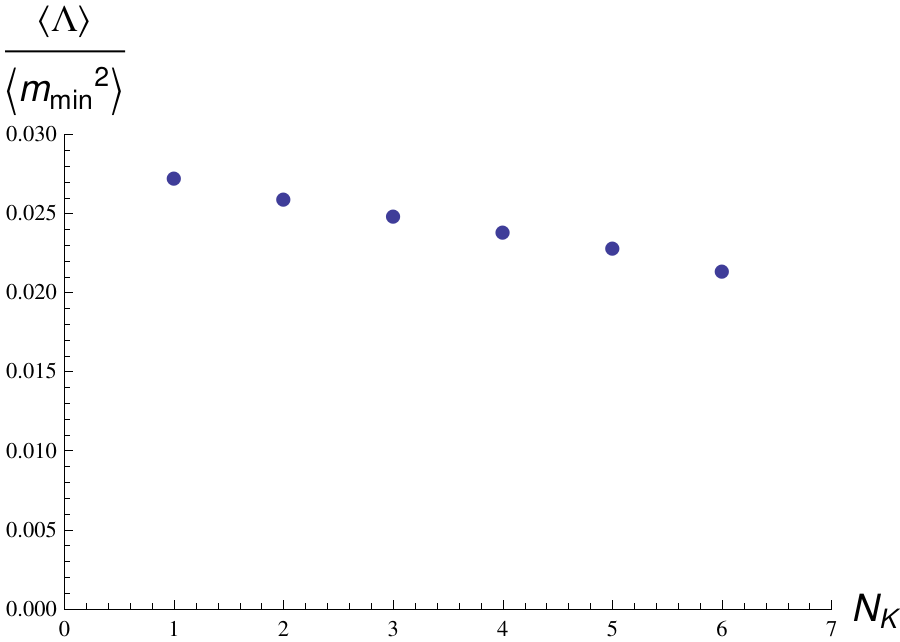}
   \end{center}
   \caption{\footnotesize The expectation value of the lowest eigenvalue of the physical mass (squared) matrix and the ratio of the expectation values.
   The estimated curve on the LHS goes as $\left< m^2_{\rm min} \right> \sim 0.0317 N_K^{-1.21} e^{-0.0422 N_K}$.
   RHS gives the  ratio $\left<  {\Lambda} \right>/\left<  {m}^2_{\rm min} \right>$ as a function of $N_K$.}
   \label{fig:mvev-uniform}
  \end{figure}

For the lowest eigenvalue $m_{\rm min}^2$ of the mass squared matrix, we observe that the expectation value $\left<m_{\rm min}^2\right>$ is decreasing as $N_K$ increases, 
as shown in Figure \ref{fig:mvev-uniform}.
We see that the ratio $\left<\Lambda\right>/\left<m_{\rm min}^2\right>$ is slowly decreasing (close to linearly) as shown in RHS of Figure \ref{fig:mvev-uniform}.

  Once the lowest eigenvalue of the mass matrix reaches small values, one may worry about the cosmological moduli problem \cite{Coughlan:1983ci,Banks:1993en,deCarlos:1993jw}, implying $m_{\rm min} \gtrsim {\cal O} (10) \, {\rm TeV} \sim 10^{-15} M_P$ to avoid the problem
  \footnote{The stringent bound for the reheating temperature $T_r \gtrsim 0.7 \, {\rm MeV}$ comes from successful neutrino thermalization and BBN \cite{Kawasaki:1999na,Kawasaki:2000en}.}.
In the current situation with uniformly distributed $W_0, A_i$, the lowest eigenvalue of mass matrix is expected to reach the the cosmological moduli bound $\left<  {m}^2_{\rm min} \right> \sim 10^{-30}$ before the current cosmological constant value $\left<  {\Lambda} \right> \sim 10^{-122}$. 
 The cosmological moduli problem depends on detail of the cosmological evolution around the era of inflation and reheating and its resolution has been proposed and studied \cite{Linde:1996cx,Takahashi:2010uw,Takahashi:2011as,Kallosh:2011qk}. 
 But even without the cosmological moduli problem, we need $N_K = 1.97 \times 10^4$ if we try to realize $\left<  {\Lambda} \right> \sim 10^{-122}$ just by the K\"ahler sector. This is probably not naturally realized in string theory compactification.

  \subsection{Sharply peaked $P(W_0)$ and uniform distribution $P(A_i)$}

  \begin{figure}[t]
   \begin{center}
    \includegraphics[width=18em]{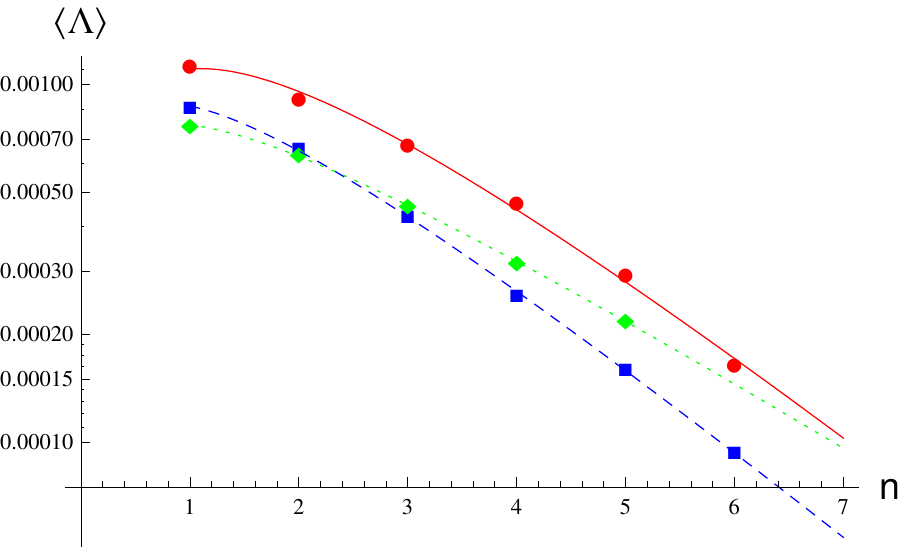}
    \includegraphics[width=18em]{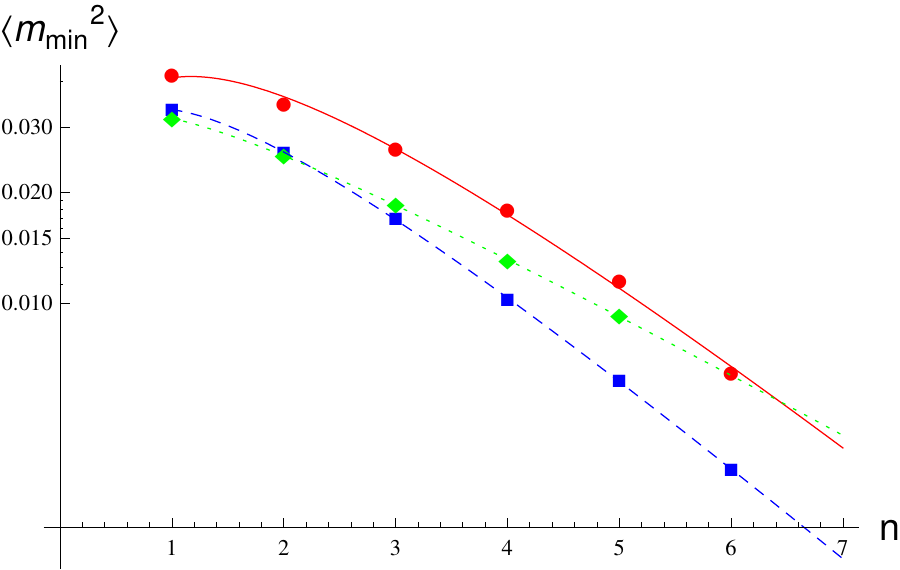}
    \includegraphics[width=18em]{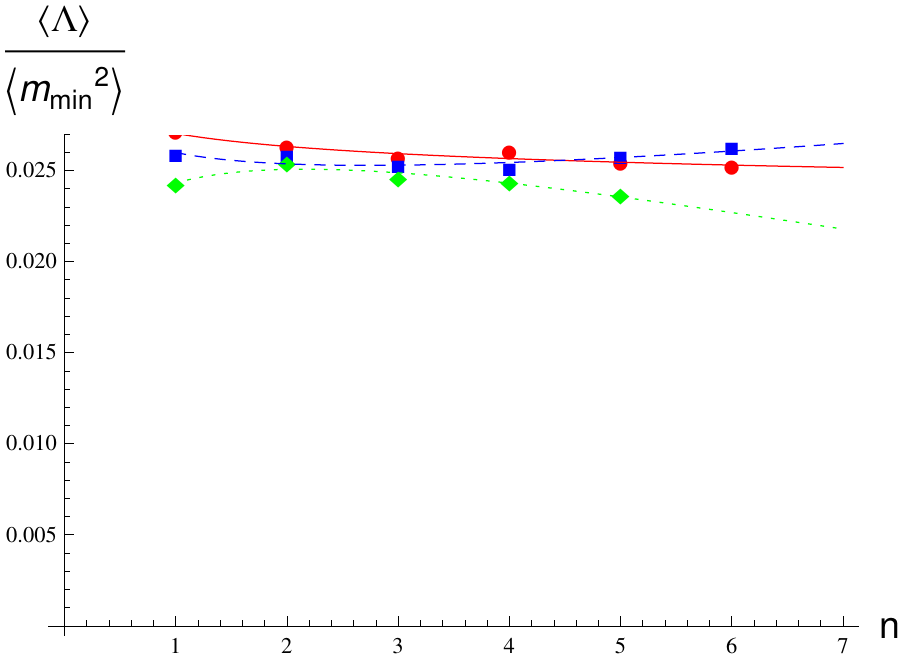}
   \end{center}
   \caption{\footnotesize The expectation values where the distribution $P(W_0)$ has a $(n-1)$ th power of logarithmic divergence at $W_0=0$, at $N_K=1$ (red, line, circle), $N_K=2$ (blue, dashed, square), and $N_K=3$ (green, dotted, diamond). Each expectation value is suppressed exponentially, though the exponential suppression at $N_K=1$ is most significant in $\left<  {\Lambda} \right>$.
   Note that the first two plots are given as log-plots.}
   \label{fig:morepeaked-W0}
  \end{figure}

  Motivated by the analysis in previous subsection, it is quite possible that the parameters $W_0, A_i$ themselves are given as products or non-trivial functions of random parameters as a result of the stabilization of the complex structure and dilaton moduli.
As a toy model, let us consider  a peaked distribution $P(W_0)$ for $W_0$
  while $A_i$ remain to have a uniform distribution as before.
  Motivated by the product distribution case (\ref{product distribution}), we simply assume 
  \begin{equation}
   P(W_0) = {1\over 15 (n-1)!} \left( \ln {15 \over |W_0|} \right)^{n-1}, \quad -15 \leq W_0 \leq 0,    
  \end{equation}
  where again the choice of sign does not affect the result.
  The number ``15'' is just a choice to increase the chance of getting $dS$ solutions in the numerical simulation.
  As we see in (\ref{eq:exponential suppression in product distribution}), the expectation value of $W_0$ is exponentially suppressed in the presence of multiple random parameters.

  Here we also assume that the values of $s, u_i$ are given as roughly of order one, and therefore negligible.
  Even though we observe that the $s, u_i$ behave differently in the simplest model for complex structure moduli stabilization in section \ref{sec:plugg-single-kahl}, it is possible that the situation will be totally different in a different class of models for complex structure moduli stabilization.
  But since we do not have enough information of the other models, here we simply ignore the direct effect of $s, u_i$
  so that we can simply extract the effect of the combination between peaked distribution of $W_0$ and K\"ahler moduli dynamics.

  In this setup, we solve the model numerically satisfying the stability constraint at positive $\Lambda$, up to $n=5$ at each $N_K=1,2,3$.
  The resultant plots are showed in Figure \ref{fig:morepeaked-W0}.
  The estimated curves for $\left<  {\Lambda} \right>$ are
  \begin{equation}
   \begin{split}
    \left<  {\Lambda} \right>_{N_K=1} =& 0.00204 n^{0.685} e^{-0.618 n},\\
    \left<  {\Lambda} \right>_{N_K=2} =& 0.00161 n^{0.474} e^{-0.617 n},\\
    \left<  {\Lambda} \right>_{N_K=3} =& 0.00124 n^{0.392} e^{-0.474 n}.
   \end{split}
  \end{equation}
  When we compare the coefficients of exponent, which is important for larger $n$, we see that the effect of the suppression is most significant around $N_K = 2$ though the increase of $n$ seems to work to make the cosmological constant smaller generically.

  On the other hand, the expectation value of minimal eigenvalue of the mass matrix are estimated by
  \begin{equation}
   \begin{split}
    \left<  {m}^2_{\rm min} \right>_{N_K=1} =& 0.0757 n^{0.725} e^{-0.619 n},\\
    \left<  {m}^2_{\rm min} \right>_{N_K=2} =& 0.0637 n^{0.548} e^{-0.644 n},\\
    \left<  {m}^2_{\rm min} \right>_{N_K=3} =& 0.0479 n^{0.260} e^{-0.413 n}.
   \end{split}
  \end{equation}
  We now see that the coefficients in exponent is quite similar to that of the one for $\left<  {\Lambda} \right>$.
  Since we have the $W_0$ dependence in the coefficient of the potential (\ref{rewritten potential for Kahler}) and this coefficient is the common part between potential and mass matrix, the similar suppression is expected even in the eigenvalue of mass matrix (see also the last plot in Figure \ref{fig:morepeaked-W0}).

  Together with the estimated functions above, now we discuss about the cosmological moduli problem.
  When we estimate the values $ {\Lambda} \sim 10^{-122}, \  {m}^2_{\rm min} \sim 10^{-30}$ with the fitted functions above, we get the numbers:
  \begin{equation}
   \begin{split}
    \begin{array}{c||c|c|c}
     N_K&1&2 &3 \\ \hline \hline
     \left<  {\Lambda} \right> \sim 10^{-122} & n\sim 451 & n\sim 449 & n\sim 583\\ \hline
     \left<  {m}^2_{\rm min} \right> \sim 10^{-30} & n\sim 113 & n \sim 107 & n\sim 163.
    \end{array}
   \end{split}
   \label{eq:n estimation of peaked W0}
  \end{equation}
  The estimated $n$ for the lower bound of moduli mass is always smaller than that for $\Lambda$.

  \subsection{Sharply peaked $P(W_0)$ and $P(A_i)$}

  In previous subsection, we considered sharply peaked $P(W_0)$ but uniform distribution for $A_i$.
  Here we consider the situation where both $P(W_0)$  and $P(A_i)$ are sharply peaked. 
  In this setup, the probability distribution functions of the parameters are given by
  \begin{equation}
   \begin{split}
    P(W_0) =& {1\over 15 (n-1)!} \left( \ln {15 \over |W_0|} \right)^{n-1}, \quad -15 \leq W_0 \leq 0,\\
    P(A_i) =& {1\over (n-1)!} \left(\ln {1\over A_i} \right)^{n-1}, \quad 0\leq A_i \leq 1.
   \end{split}
  \end{equation}

  \begin{figure}[t]
   \begin{center}
    \includegraphics[width=18em]{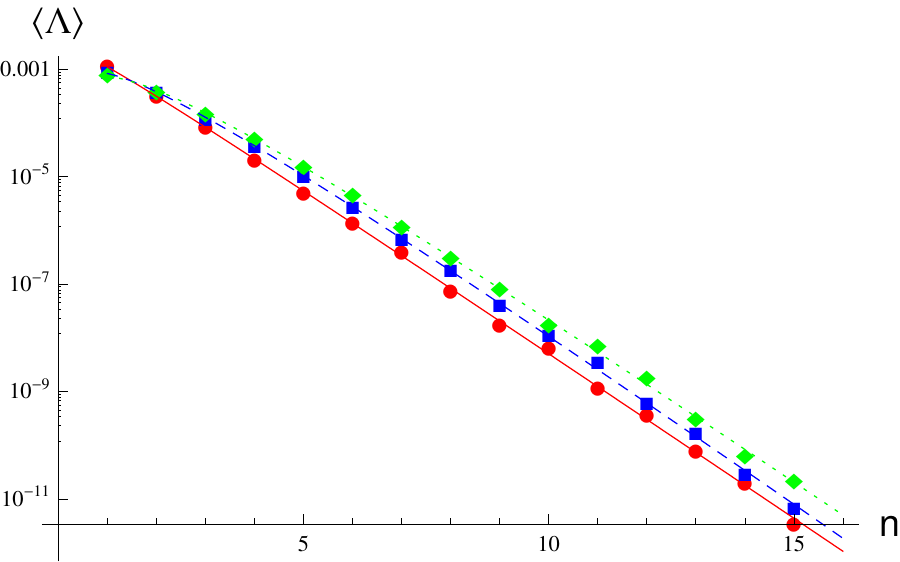}
    \includegraphics[width=18em]{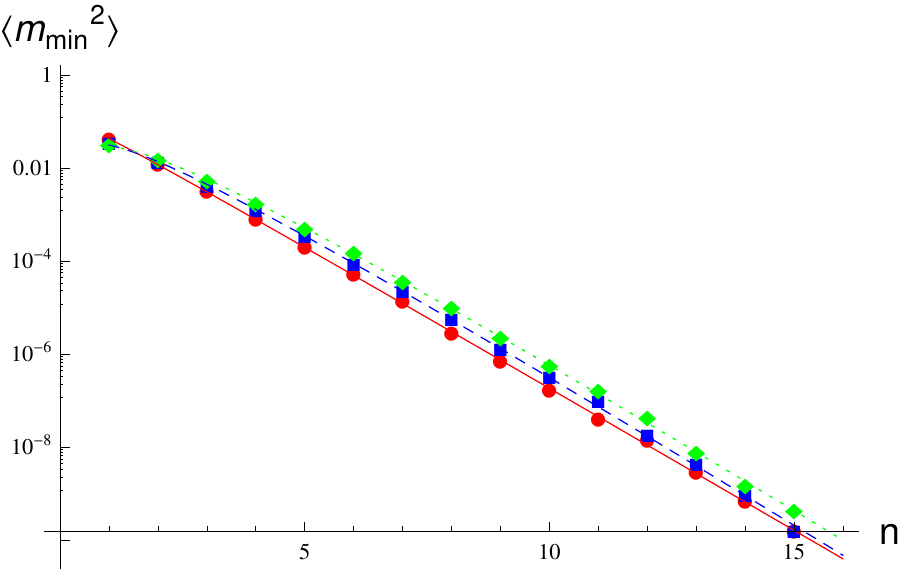}
    \includegraphics[width=18em]{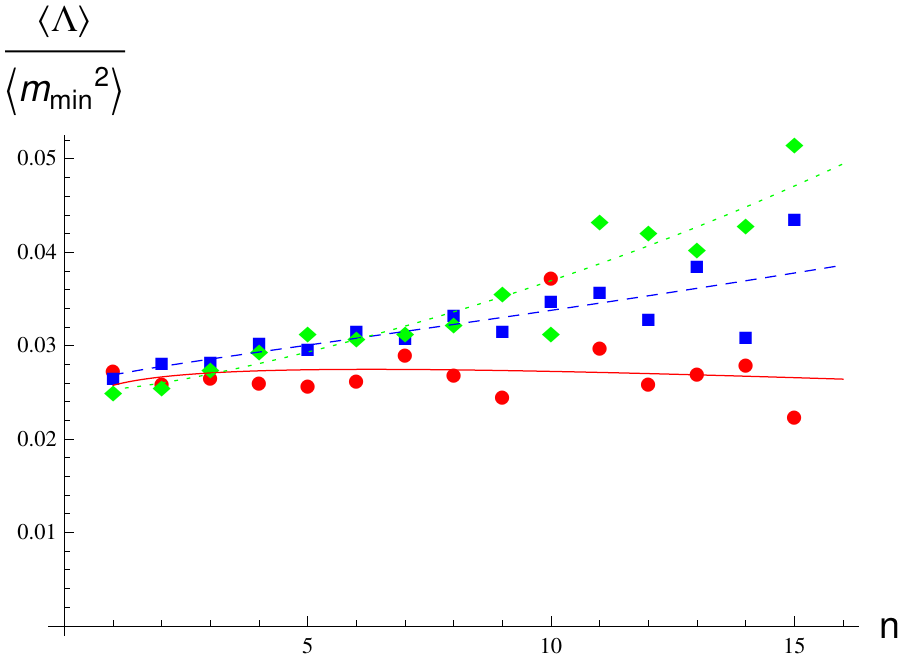}
   \end{center}
   \caption{\footnotesize The expectation values with peaked distributions $P(W_0)$ and $P(A_i)$ for the parameters $W_0, A_i$, with $(n-1)$ th power of logarithmic divergence at $W_0=0$ and $A_i=0$ respectively. The expectation values are given for $N_K=1$ (red, line, circle), $N_K=2$ (blue, dashed, square), and $N_K=3$ (green, dotted, diamond). 
   Each expectation value is suppressed exponentially, though the exponential suppression at $N_K=1$ is most significant in $\left<  {\Lambda} \right>$.
   Notice that the first two plots are given as log-plots.}
   \label{fig:morepeaked-W0Ai}
  \end{figure}

  In the numerical calculation, we can get a result even at $n=15$ in this setup.
  This is probably related to the preference of small values of $B_i = {A_i /A_1}$ to stabilize the model at positive $\Lambda$.
  Now since the distribution of $A_i$ prefers smaller values, the resultant $B_i$ are also likely smaller than that in case of uniform distribution.

  The plots for the expectation values are given in Figure \ref{fig:morepeaked-W0Ai}.
  The estimated functions for $\left<  {\Lambda} \right>$ here are given by
  \begin{equation}
   \begin{split}
    \left<  {\Lambda} \right>_{N_K=1} =& 0.00462 n^{0.252} e^{-1.43 n},\\
    \left<  {\Lambda} \right>_{N_K=2} =& 0.00384 n^{1.07} e^{-1.53 n},\\
    \left<  {\Lambda} \right>_{N_K=3} =& 0.00328 n^{1.31} e^{-1.49 n}.
   \end{split}
  \end{equation}
  It is interesting that each coefficient of the exponent is below $- 2 \ln 2 \sim - 1.39$.
  Since the coefficient appearing in (\ref{rewritten potential for Kahler}) is given as a product of $A_1 W_0$, we may naively expect an exponential suppression of the form $e^{-2 n\ln 2}$.

  The estimated functions for the lowest eigenvalue of mass matrix become
  \begin{equation}
   \begin{split}
    \left<  {m}^2_{\rm min} \right>_{N_K=1} =& 0.178 n^{0.188} e^{-1.40 n},\\
    \left<  {m}^2_{\rm min} \right>_{N_K=2} =& 0.146 n^{1.05} e^{-1.55 n},\\
    \left<  {m}^2_{\rm min} \right>_{N_K=3} =& 0.136 n^{1.35} e^{-1.55 n}.
   \end{split}
  \end{equation}
  Again, each coefficient of the exponent is below $-2 \ln 2 \sim -1.39$.
  The non-trivial complicated moduli couplings push the expectation value down further in addition to the suppression by the coefficient.
  But here, as in the second plot of Figure \ref{fig:morepeaked-W0Ai}, the expectation value at larger $N_K$ with fixed $n$ always stays at lower value, meaning that the off-diagonal components in the Hessian tends to push the lowest eigenvalue further down as the rank $N_K$ of the Hessian increases. This effect tends to push up the ratio $\left<  {\Lambda} \right>/\left<  {m}^2_{\rm min} \right>$.

  Now we are ready to consider the cosmological moduli problem.
\begin{equation}
   \begin{split}
    \begin{array}{c||c|c|c}
     N_K&1&2 &3 \\ \hline \hline
     \left<  {\Lambda} \right> \sim 10^{-122} & n\sim 194 & n\sim 184 & n\sim 189\\ \hline
     \left<  {m}^2_{\rm min} \right> \sim 10^{-30} & n\sim 48 & n \sim 46 & n\sim 47.
    \end{array}
   \end{split}
   \label{eq:n estimation of peaked W0Ai}
  \end{equation}
  We see that the required number of $n$ for the lower bound of mass is always smaller than that for the cosmological constant $\Lambda$.

  If we compare the result here with the one in previous subsection (\ref{eq:n estimation of peaked W0}), we see that the required $n$ here is much smaller.
  This is nothing but a benefit of the additional peakiness in $A_i$, which is considered to come from the effect of multiple moduli stabilization in the complex structure sector.
  Although so far we have considered up to $N_K = 3$ in this subsection just for simplicity, it is quite possible that we can explain {\it at least} the small value of $\Lambda$ naturally in terms of the variety of string landscape.

\section{Discussions and Remarks \label{sec:discussions-remarks-}}

When there is no preference for a small cosmological constant $\Lambda$, extreme fine-tuning is necessary to end up with an exponentially small $\Lambda$, when the Planck mass $M_P$ is the only scale in the model. Here we show that simple properties of probability theory can lead to a vanishingly small 
$\Lambda$ in some stringy scenarios.
In particular, the peaking of the probability distribution $P(\Lambda)$ at $\Lambda=0$  is quite robust.
Admittedly, the particular model we study is chosen more for its solvability than for its phenomenological properties. Nevertheless, it is a model that have been extensively studied in the literature and is non-trivial enough to present some interesting features to serve our specific purpose. This possible stringy mechanism for the preference of an exponentially small $\Lambda$ should be further explored in more realistic stringy scenarios.

Within the model studied in this paper, different ways in implementing the distributions of the flux parameters lead to rather different behaviors of $\Lambda$ as a function of $h^{2,1}$. This is because the effect of the distribution of the flux parameters get amplified by a large $h^{2,1}$. A better understanding of the stringy dynamics will be very useful in finding the actual distributions of the flux parameters.

In this paper, we consider the simplest model for the complex structure moduli stabilization, in which we only have linear terms of the complex structure moduli $U_i$ in the superpotential  $W_0$.
We can expect more peaked distributions in $w_0$ if there are cross couplings among $U_i$ in the superpotential.
Actually, such cross couplings are quite ubiquitous even in toroidal orientifolds \cite{Lust:2005dy}.
Therefore it would be important to investigate the probability distributions $P(w_0)$ and $P(\Lambda)$ in the more realistic  models to see whether and under what conditions $\Lambda$ will be naturally vanishingly small. 
We are encouraged by a preliminary investigation where the peaking behavior of $P(w_0)$ seems to strengthen when quadratic terms of complex structure moduli are introduced into the superpotential. Obviously a more detailed study is required to get a full picture of the distribution of $\Lambda$.

In this paper, we assume all parameters are physical parameters that include radiative corrections. In this sense, we are calculating $P(\Lambda)$ for the physical $\Lambda$. We are interested in the situation that the corrections are small compared to the original inputs and therefore can be effectively absorbed in the parameters assigned in the model, which we randomize.
If we let all the parameters in the model be bare parameters, then the calculated $P(\Lambda)$ is obviously for the bare $\Lambda$. In general, when the parameters are randomized, the ranges and distributions of the bare parameters and that of the physical parameters will be different. Since we have only a vague notion of their ranges and distributions, it is important to know when their ranges and distributions will make a difference in the resulting physics.
As we believe, the peaking behavior of $P(w_0)$ and $P(\Lambda)$ are quite generic for reasonable ranges and distributions, while $\left< |\Lambda| \right>$ is more sensitive to the details.

Even if the radiative and other corrections contribute significantly to the potential, we expect that the peaking behavior is still there since the correction terms will bring in non-trivial moduli dependence and the resulting correlation most likely strengthens the peaking behavior in general.
It is interesting to see if how the peaking behavior is affected by the presence of the other types of $\alpha'$ and string loop corrections.
Also the backreaction of the $\alpha'$-correction and non-perturbative terms generating the potential for K\"ahler moduli would be important for complex moduli sector.

Recently it is argued that the D7 tadpole cancellation condition \cite{Collinucci:2008pf} of the system and the holomorphicity of D7-branes in flux compactification restricts the maximal rank of gauge group on D7-branes \cite{Cicoli:2011qg,Louis:2012nb}, which is a candidate to derive non-perturbative terms as a result of the gaugino condensation.
The explicit solution for $dS$ space was obtained with ${\mathbb C P}_{1,1,1,6,9}^4$ even in the presence of the constraint for $a_i$ \cite{Louis:2012nb}.
Although we have analyzed statistical behavior only by randomizing $W_0, A_i$ with the other parameters fixed just for simplicity in this paper, it would be interesting to see the effect by randomizing the parameter $a_i$ within the known restrictions/constraints.

The very small mass of the lightest modulus may pose the cosmological moduli problem \cite{Coughlan:1983ci,Banks:1993en,deCarlos:1993jw}. It has been proposed that this problem may be avoided in some senarios \cite{Takahashi:2010uw,Takahashi:2011as,Kallosh:2011qk} based on \cite{Linde:1996cx}.  For example,  a coupling generated by a strong dynamics helps to suppress exponentially the moduli coherent oscillation itself. This mechanism requires multi-field inflation to avoid the eta-problem and it turns out that the multi-field inflation is rather ubiquitous in string compactification. Another possibility is thermal inflation. This scenario is totally compatible with observations and helps to dilute the light moduli energy density, which may be over-produced otherwise.  Whether the proposed resolution within supergravity applies to Type IIB string theory remains to be seen. Clearly a better understanding of this issue will involve the physics in the early universe. This paper is focused on the probabilistic property of the mathematical functional form of today's $\Lambda$. We may impose the resolution to the light moduli problem as a constraint on model building for the early universe after inflation. 

It is important to point out that the effective action for $N$ scalar fields takes the general form
\begin{equation}
 \Gamma = \int d^4x \left(-V(\phi_i) + Z_{jk}(\phi_i) \partial \phi_j \partial \phi_k +  \cdot \cdot \cdot \right). \notag
\end{equation}
In perturbative theory, $Z_{jk} \simeq \delta_{jk}$. For a large matrix $Z_{jk}$, small off-diagonal terms can lead to non-trivial (even random) behaviors \cite{Chen:2011ac}.  For classical stability, positivity of the Hessian $\partial_j \partial_k V$ is enough. However, the mass eigenvalues of the scalar fields strongly depends on $Z_{jk}$ as well. It will be interesting to see what happens to the lightest scalar modes (both their masses and their couplings to matter fields) in a more realistic string model.

The tunneling probability from the meta-stable vacuum to the supersymmetric decompactified vacuum is, using the Hawking-Moss formula,
\begin{equation}
{\cal P} = \exp \left( - 24\pi^2 \left[ \frac{1}{\Lambda} - \frac{1}{V_H} \right] \right). \notag
\end{equation}
For an exponentially small $\Lambda$, we find that typical barrier height easily has $V_H \gg \Lambda$;
for all practical purposes, de-compactification simply does not happen, as in \cite{Kachru:2003aw}. So we are quite safe in this respect.

Naively, we have supersymmetric breaking scale $M_{SUSY} > 1$ TeV, while $\Lambda \sim M_{SUSY}^4$. These 2 properties are clearly phenomenologically incompatible. To resolve this discrepancy in our scenario, $\Lambda$ should drop exponentially as the number of complex structure moduli increases while the soft supersymmetry breaking mass $m_{soft}$ drops slower. These behaviors crucially depend both on the functional forms of $\Lambda$ and $m_{soft}$ on the flux parameters and on the distributions of the flux parameters.

\section*{Acknowledgment}

We have benefited from discussions with Lam Hui, Liam McAllister, Fernando Quevedo, Markus Rummel, Gary Shiu, Zheng Sun, Alexander Westphal, Timm Wrase and Tsutomu Yanagida.
YS is grateful to Isaac Newton Institute for Mathematical Sciences and the organizers of ``String Phenomenology 2012'', and also the organizers of ``The 3rd UTQuest workshop ExDiP 2012 on Superstring and Cosmophysics'' for their great hospitality, where some of the results were presented.

\appendix

 \section{Some toy models \label{toy1}}

  To get some feeling on the peaking property for the probability distribution $P(\Lambda)$ of $\Lambda$,
consider the single field polynomial model. At least three terms with different powers are required to achieve the metastable vacua with a positive $\Lambda$ \cite{Maloney:2002rr,Silverstein:2004id,Silverstein:2007ac,Haque:2008jz}.
Since we are interested in only the meta-stable solution, we may consider a model of a simple minimal form:
  \begin{equation}
  \label{A11}
V(\phi) = a \phi - {b \over 2} \phi^2 + {c \over 3!} \phi^3
 \end{equation}
 where all parameters are positively defined for simplicity,
 and have the scale $M_P^4$.
Now, we impose the stability $\partial_\phi^2 V|_{\rm min} >0$ at the extremal points given by $\partial_\phi V=0$.
These conditions imply
\begin{equation}
 \Delta \equiv \sqrt{b^2-2 ac} >0, \quad \phi_{\rm min}= {b + \Delta \over c}.
\end{equation}
Plugging back to the potential, we get
\begin{equation}
\Lambda\equiv V_{\rm min}= \frac{(b+\Delta)^2(b-2\Delta)}{6 c^2}.
\label{A12}
\end{equation}
So we see that the sum form (\ref{A11}) is converted to the product form (\ref{A12}).

If we impose the positivity of the potential $\Lambda \geq 0$, then the parameter region is further restricted by
\begin{equation}
 1< C_p \leq {4\over 3}, \quad C_p \equiv {b^2 \over 2 a c},
\end{equation}
where the lower bound is given by the stability while the upper bound is for the positivity.
Whenever we satisfy the region of the combined parameter above, there exists a minimum with positive $\Lambda$.
Since we are interested in the region with small $\Lambda$s, let us further focus on the region around $C_p \sim 4/3$.
Then the potential becomes
\begin{equation}
 \Lambda \sim {9\over 8} {a b \over c} \left( {4\over 3} - C_p \right).
  \label{eq:approximate Lambda for polynomial}
\end{equation}

\begin{figure}[t]
 \begin{center}
  \includegraphics[width=18em]{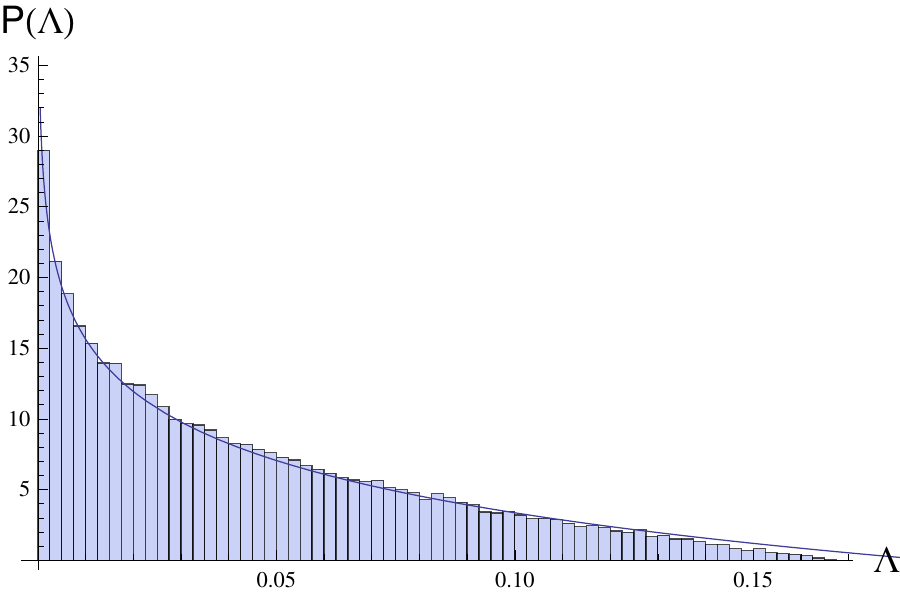}
  \includegraphics[width=18em]{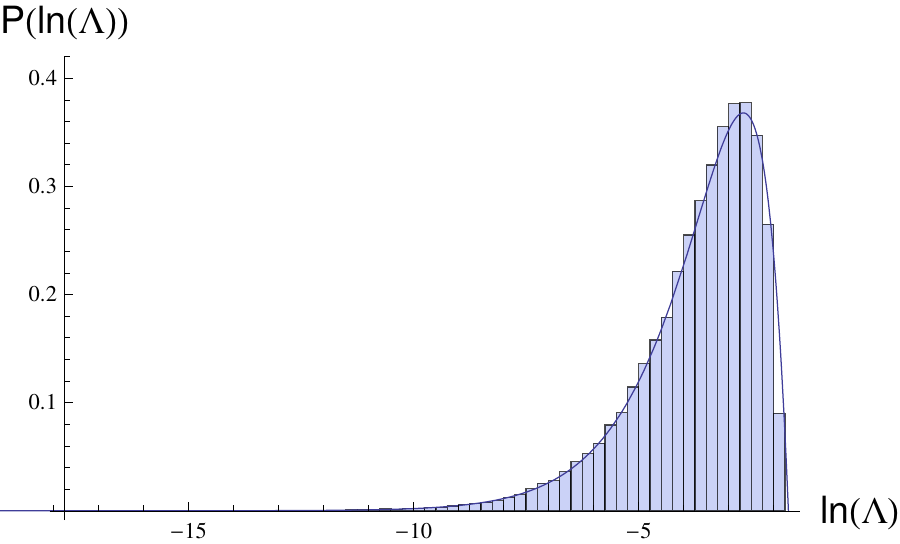}
 \end{center}
 \caption{\footnotesize The comparison of the analytical (the curve) and the numerical (the histogram) distributions $P(\Lambda)$ vs $\Lambda$ (LHS) and $P(\ln  {\Lambda})$ vs $\ln  {\Lambda}$ (RHS) in the polynomial model (\ref{A11}).}
 \label{fig:polynomial}
\end{figure}

Next we analyze the distribution of the approximate minima (\ref{eq:approximate Lambda for polynomial}).
To make the argument simple, we here fix $c=1$ and take $a,b$ as random parameters obeying uniform distribution within a range $0\leq a,b \leq 1$.
Then the probability distribution function 
is estimated to be
\begin{equation}
 P(\Lambda) = N_0^{-1} \int dC_p \int_0^1 d {a} d {b} \, \delta\left({9\over 8} { {a}  {b}} \left({4\over 3} - C_p \right)-  {\Lambda} \right)\, \delta\left({ {b}^2 \over 2  {a}} - C_p \right),
\end{equation}
where $N_0$ is the normalization constant.
Performing the last two integrals for the two $\delta$-functions, being careful with the integration regions, we get
\begin{equation}
 P( {\Lambda}) = N_0^{-1} \int dC_p\,  {1 \over 3 C_p (4/3-C_p)}, \quad
  1< C_p \leq {12 \over 9 + 16  {\Lambda}}.
\end{equation}
The constraint for $C_p$ also suggests $0 \leq  {\Lambda} < 3/16$.
The normalization constant $N_0$ is calculated such that $\int d {\Lambda} P( {\Lambda}) = 1$, which gives $N_0 = 1/24$, so
\begin{equation}
 P( {\Lambda}) = {16 \over 3} \ln \left[{3 \over 16  {\Lambda} }\right].
  \label{eq:divergent PDF in polynomial}
\end{equation}
where $P( {\Lambda})$ is divergent at ${\Lambda} = 0$.
This is nothing but the outcome of the product form of the coefficient obtained in (\ref{eq:approximate Lambda for polynomial}).
If we allow the parameter $c$ to have a distribution (even including zero), the divergence at ${\Lambda} = 0$ is still present (see the argument for ratio distributions, e.g. in \cite{Sumitomo:2012wa}).
However, $P( {\Lambda})$ will have a longer tail at large $\Lambda$.

Although we have analyzed the approximate potential (\ref{eq:approximate Lambda for polynomial}), we can easily check  numerically the divergent behavior (\ref{eq:divergent PDF in polynomial}) at the minimum (\ref{A12}) in the full form of the potential.
In Figure \ref{fig:polynomial}, the analytical function (\ref{eq:divergent PDF in polynomial}) perfectly agrees with the numerical histogram especially at lower values, given uniformly distributed random values for $0\leq  {a}, {b} \leq 1$ while fixing $ {c}=1$.

Next, consider a simple multi-field case, 
$$V(\phi_i) = A_i\phi_i + \frac{1}{2} B_{ij} \phi_i \phi_j $$ 
We see that $$V_{\rm min}= - \frac{1}{2} A^TB^{-1}A$$ which is a sum of terms for the multi-field cases. Clearly the $B_{ij}$ coupling alone is not enough for $V_{\rm min}$ to take the product form. Higher couplings for tighter interactions among the fields are necessary to bring $V_{\rm min}$  to the product form.

Although we do not have an analytic form for $V_{\rm min}$ for more complicated $V(\phi)$, it is easy to convince oneself (and check numerically) that a divergent $P(\Lambda)$ at  $\Lambda =0$ for meta-stable $V_{\rm min}$ is quite typical for $V$ in which all terms are tightly coupled to each other.

For more complicated $V(\phi)$, $V_{\rm min}$ takes a more complicated form. We are looking for 2 features :
(1) whether the probability distribution $P(\Lambda)$ has a peaking (maybe even divergent) behavior at $\Lambda=0$, and the peakiness becomes more pronounced as the number of fields and their couplings (parameters treated as random variables) increase; and (2) whether the expectation value of the magnitude $\left< \Lambda  \right>$ is dropping exponentially fast as the number of fields increases.

\section{$P(\Lambda)$ in the single K\"ahler modulus case \label{sec:prob-distr-plambda}}

We show here the way to estimate the probability distribution $P(\Lambda)$ in the single K\"ahler modulus case discussed in section \ref{sec:revi-analyt-study}.
Let us define
\begin{equation}
 w_1 = - W_0, \quad w_2 = A_1,  \quad c = {w_1 \over w_2}={64 \sqrt{2} \over 27}C.
\end{equation}
and set the value $a_1=\gamma_1=  \hat{\xi}=1$.
Neglecting the overall coefficient, the model is  simplified to be
\begin{equation}
 \begin{split}
  z =  w_1 w_2 (c-c_0), \quad c_0 \leq c= {w_1 \over w_2} < c_1,
 \end{split}
 \label{eq:simplified z}
\end{equation}
where $c_0 \sim 12.2$ and $c_1 \sim 13.0$.
The cosmological constant is related with this $z$ through the relation $ {\Lambda} = {3 z / (2500 \sqrt{5})}$.

 Now the probability distribution of $z$ is determined by
 \begin{equation}
 \begin{split}
  P(z) = N_0^{-1} \int_{c_0}^{c_1} dc \int_0^1 dw_1 dw_2 \, \delta \left(w_1 w_2 (c-c_0) \right) \, \delta \left({w_1 \over w_2} - c\right),
 \end{split}
 \end{equation}
where $N_0$ is the normalization constant.
Since this is the constrained integrations by two delta functions, the integrated region is further constrained.
The last two integrations can be achieved easily and we get
\begin{equation}
 \begin{split}
  P(z) = \left\{
  \begin{array}{l}
   N_0^{-1} \int dc \, {1\over 2c(c-c_0)} \quad  {\rm for} \ 0\leq z \leq c(c-c_0), \ c\leq 1, \\
   N_0^{-1} \int dc \, {1\over 2c(c-c_0)} \quad  {\rm for} \ 0\leq z \leq {c-c_0 \over c}, \ c\geq 1.
  \end{array}\right.
 \end{split}
\end{equation}
Since the region of $c$ of interest is $1<c_0 \leq c < c_1$, we can integrate over $c$ and
\begin{equation}
 \begin{split}
 P(z) =& N_0^{-1} \int_{c_0 \over 1-z}^{c_1} dc {1\over 2c(c-c_0)}
 = {c_1 \over c_1 - c_0} \ln \left[c_1 - c_0 \over c_1 z \right],\\
  N_0 =& \int_0^{c_1 - c_0 \over c_1} dz \int_{c_0 \over 1-z}^{c_1} dc {1\over 2c(c-c_0)} = {c_1 - c_0 \over 2 c_1 c_0}.
 \end{split}
\end{equation}
The probability distribution of $ {\Lambda}$ can be estimated via the relation $P( {\Lambda})\, d {\Lambda} = P(z)\, dz$ by,
\begin{equation}
 P( {\Lambda}) = {2500 \sqrt{5} \over 3} {c_1 \over c_1 - c_0} \ln \left[{3 \over 2500 \sqrt{5}} {c_1 - c_0 \over c_1  {\Lambda} }\right].
  \label{eq:PDF for Lambda1}
\end{equation}
which is (\ref{eq:PDF for Lambda}).

 \section{Special symmetric case for $s$ and $w_0$ \label{sec:discussion-solving-s}}

Some of the manifolds have special symmetries so it is interesting to see the impact of them on the value of $s$ and the probability distribution $P(w_0)$.
  
 Let $r_i=d_i/b_i$ and $r_0=c_2/c_1$.
  Let us consider the special situation,
  \begin{equation}
   r \equiv r_1=r_2= ... =r_i=...=r_{h^{2,1}},
  \end{equation}
  while $r \ne r_0$.
  Following  from (\ref{mc4}), all complex structure moduli $u_i$ are equal.
  Then the solution of (\ref{mc7}) is given by
  \begin{equation}
   s={1\over 2r r_0}\left[({h^{2,1}}-1)(r- r_0) \pm \sqrt{({h^{2,1}}-1)^2(r-r_0)^2 +4r r_0}\right],
  \end{equation}
  which, for large ${h^{2,1}}$ yields, for $r>r_0$,
  \begin{equation}
   s \sim ({h^{2,1}}-1){r-r_0 \over  rr_0}, \quad {\rm or} \quad {1\over (h^{2,1}-1)(r-r_0)},
    \label{eq:special sol for s at large h^2,1}
  \end{equation}
 in which case,
 $r$ must be close to $r_0$ and
 $r_0>r$ for $s$ to be large and positive.
In this case, we now have
 \begin{equation}
 \label{aawm}
W_0|_{\rm min} = \frac{c_1}{{h^{2,1}}} (1+sr_0)\left(\frac{1-sr}{1+sr} \right).
  \end{equation}
It is easy to see that the strong peaking behavior of $P(w_0)$ is absent here.

At the symmetric point, $b_i=d_i$ or $r=r_i=1$, (\ref{eq:special sol for s at large h^2,1}) reduces to 
\begin{equation}
 \label{mc9}
  s \sim ({h^{2,1}}-1) \left(\frac{c_1}{c_2} -1\right).
\end{equation}
 Recall that we like $s$ to be large and positive, say $s \approx 10$, so we need $c_1 >c_2$.

The following picture emerges. As we move towards the symmetric points, the probability distribution $P(w_0)$ loses its peaking behavior at $\Lambda=0$. This is simply because (the random) parameter space collapses at the symmetric points.

  \section{Stabilization in two and three K\"ahler moduli models \label{sec:stab-two-three}}

  For the analysis of the potential for multi-K\"ahler moduli in section \ref{sec:numer-analys-distr}, we elaborate on the stability analysis in the cases of two and three K\"ahler moduli.
  Here we just focus on the bracket part of (\ref{rewritten potential for Kahler}) since the coefficient part does not touch the stability.
  For convenience, we further define
  \begin{equation}
   \begin{split}
    {\cal U} \equiv  {2 C \over 9 (x_1^{3/2} - \sum_{j=2} \delta_j x_j^{3/2} )^3} - {x_1 e^{-x_1} \over (x_1^{3/2} - \sum_{j=2} \delta_j x_j^{3/2} )^2} - \sum_{i=2} {B_i x_i e^{-x_i}\over (x_1^{3/2} - \sum_{j=2} \delta_j x_j^{3/2} )^2}.
   \end{split}
   \label{bracket part of the potential}
  \end{equation}

  \subsection{Two K\"ahler moduli}
  
  Let us start with the two moduli case.
  In the region $\delta_2 \ll 1$, the stabilization of $x_1$ is little changed from that of the single K\"ahler case.
  Therefore the moduli stabilization for $x_2$ requires small $B_2$ so the third term in (\ref{rewritten potential for Kahler}) is comparable with the other terms proportional to $\delta_2$.
  The situation begins to differ from the single K\"ahler case as $\delta_2$  increases.
  But since the volume is defined to be positive and we expect the fixed moduli value of $x_2$ to be comparable with $x_1$, $\delta_2$ should not be too large.
  So hereafter we take $\delta_2=1$ just for simplicity.
  For stability, we need to be careful about the positivity of mass matrix.
  According to {\it Sylvester criteria} (see e.g. \cite{Gilbert:1991}, which is also used in \cite{Shiu:2011zt}), we consider positivity of $\partial_{x_1}^2 {\cal U}$ and the determinant of the $2 \times 2$ matrix $\partial_{x_i} \partial_{x_j} {\cal U}$ , simultaneously with the positivity of the extremal points.

  \begin{figure}[t]
   \begin{center}
    \includegraphics[width=13em]{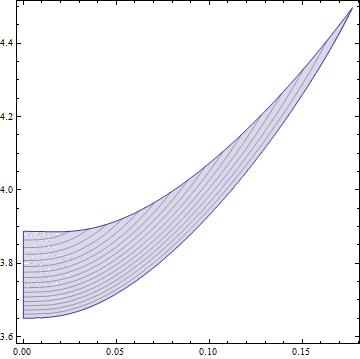}
   \end{center}
   \caption{\footnotesize The validity region in $B_2-C$ plane for meta-stable $dS$ vacua with two moduli. The $x$-axis is $B_2$ and the $y$-axis is $C$.
   The contours inside the region describe level lines of ${\cal U}$, not the magnitude of $\Lambda$.}
   \label{fig:stable-region}
  \end{figure}

  The extremal conditions $\partial_{x_i} {\cal U}=0$ are solved by
  \begin{equation}
   \begin{split}
    C =&  {e^{-x_1} \over \sqrt{x_1}(x_2-1)} \\
    &\times \left[(2x_1^3 + x_1^4)(x_2-1) - x_1^{3/2} x_2^{3/2} (x_2 -4) + (x_1 x_2^3 - x_2^3) (x_2+2) - x_1^{5/2} x_2^{3/2} (2 x_2 + 1) \right],\\
    B_2 =& {e^{-x_1 + x_2} (x_1-1)  \over (1-x_2)} \sqrt{x_2 \over x_1}.
   \end{split}
   \label{extremal conditins in two Kahler space}
  \end{equation}
  If we take $x_2 \rightarrow 0$, the conditions recover that for the single K\"ahler case.
  Together with the positive mass squared conditions above, Figure \ref{fig:stable-region} shows the allowed parameters region with respect to $C, B_2$ for meta-stable $dS$ vacua.
  In the system above, the positivity constraint of the determinant controls the upper bound of the parameter region in Figure \ref{fig:stable-region}, while the lower bound is given by the positivity constraint ${\cal U}>0$.
  The contours inside the region describe level lines of ${\cal U}$, not the magnitude of $\Lambda$. On the other hand, $\Lambda=0$ at the top boundary curve, and $\Lambda$ increases as we move down away from this boundary.
  The numerical values of the parameters in the region are roughly given by
  \begin{equation}
   \begin{split}
    3.650 \lesssim C& \lesssim 4.498,\quad    0 < B_2 \lesssim 0.1769.
   \end{split}
   \label{constraints two moduli}
  \end{equation}
  For instance, the lower bound of the parameters requires the potential minimum at
  \begin{equation}
   x_1 = 2.5, \quad x_2 = 0,
  \end{equation}
  while the upper bound of the parameters suggests the minimum at around
  \begin{equation}
   x_1 \sim 3.881, \quad x_2 \sim 0.7085.
  \end{equation}
  Note that the value of ${\cal U}$ increases gradually with $C$, similar to the single K\"ahler modulus case.

  Note that the validity range of the $C$ parameter is different from the one obtained for single K\"ahler moduli.
  The result for single K\"ahler modulus case is \cite{Rummel:2011cd} 
  \begin{equation}
   3.650 \lesssim C \lesssim 3.887, \quad B_2 =0.
  \end{equation}
  Therefore we see that the additional modulus dependence alleviates the constraint for $C$ parameter, but still restricts $B_2$ small.
  We may understand the smallness of $B_2$ as a consequence of the positivity of the volume ${\cal V}$ defined in (\ref{effective potential for multi Kahler}).
  The required smallness of $\sum_{i=2} \gamma_i t_i/ \gamma_1 t_1$ affects the combinations $A_i/A_1$ to stabilize the potential in entire K\"ahler moduli space.
  Therefore $A_i/A_1$ cannot be large and are about the same order of $x_i/x_1$ (or the combination of $\delta_i x_i/x$) at the $dS$ minima.

 \subsection{Three K\"ahler moduli}
 
 Now let us proceed to the model with three K\"ahler moduli case.
 Since it is difficult to draw figures with continuous surfaces, here we use the numerical data given random values for parameters assigned to the model.
 Similar to the two moduli case, we set $\delta_2 = \delta_3 = 1$ just for simplicity.

\begin{figure}[t]
  \begin{center}
   \includegraphics[width=12em]{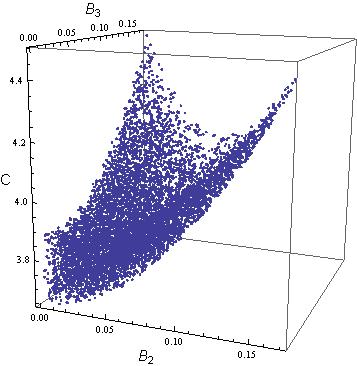}
   \includegraphics[width=14em]{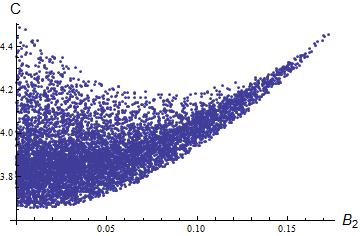}
   \includegraphics[width=14em]{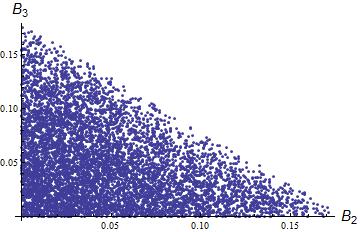}
  \end{center}
  \caption{\footnotesize The parameter region in the $B_2-C-B_3$ 
 plot shows the stable minima at positive $\Lambda$ with three moduli. 
  The last two figures display the data on a projected plane.}
  \label{fig:3kahler-parameters}
 \end{figure}

 Given the values for parameters $C, B_2, B_3$ obeying uniform distribution, we get the region of parameter sets which satisfy the stability condition at positive $\Lambda$, shown in Figure \ref{fig:3kahler-parameters}.
 All points here satisfy the stability constraint at extremal points with $\Lambda \ge 0$.
 We see that the stable parameter region for three moduli case includes the stable region for two moduli case as a boundary. 
 The restricted ranges for the parameters are given by
 \begin{equation}
  3.650 \lesssim C \lesssim 4.498, \quad 0 < B_2 \lesssim 0.1769, \quad 0 < B_3 \lesssim 0.1769.
  \label{DD7}
 \end{equation}
 The parameters are correlated so they are further constrained.

 \subsection{Checking the validity of the approximate potential}
 
 So far we have seen what happens in the approximate potential (\ref{bracket part of the potential}). Here let us compare the resultant parameter space with that of the full potential (\ref{effective potential for multi Kahler}).
 The full potential is no longer written down in terms of the combined parameters (\ref{eq:combined parameters}).
 Therefore we give values for $W_0, A_i$ and set $a_i = \gamma_i = \hat{\xi} = 1$ for simplicity, and solve for moduli fields $t_i$.

 \begin{figure}[t]
  \begin{center}
   \includegraphics[width=18em]{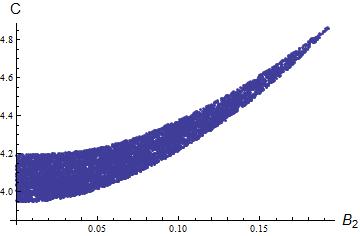}
   \includegraphics[width=15em]{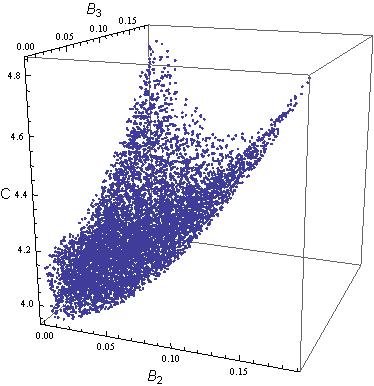}
  \end{center}
  \caption{\footnotesize The stable parameter region for dS vacua in the full potential (\ref{effective potential for multi Kahler})  for the two and three moduli case.}
  \label{fig:valid-parameters-fullpotential}
 \end{figure}

 The full potential for two K\"ahler moduli suggests the stable parameter region for $C, B_2$ as in LHS of Figure \ref{fig:valid-parameters-fullpotential}, where 
 \begin{equation}
  3.95 \lesssim C \lesssim 4.87, \quad 0 < B_2 \lesssim 0.193.
 \end{equation}
Although the boundary values for parameter region are changed a little from the approximate ones (\ref{constraints two moduli}) for the approximate potential (\ref{rewritten potential for Kahler}), the essential feature is unchanged, including the shape of the region which is crucial for the distribution of $\Lambda$.

 In the three moduli model, the full potential also suggests the similar parameter region for stability to that for the approximate potential.
The stability condition at positive $\Lambda$ is satisfied if the parameters are within the region in RHS of Figure \ref{fig:valid-parameters-fullpotential}, suggesting roughly
 \begin{equation}
  3.95 \lesssim C \lesssim 4.87, \quad 0 < B_2 \lesssim 0.193, \quad 0 < B_3 \lesssim 0.193.
 \end{equation}
 Again, though the values are changed a little from (\ref{DD7}), the shape of stable region is the same as before.

 From the examples of the analysis of the full potential above, we see the basic agreements between the approximate potential and the full potential, at least for the distribution analysis of small $\Lambda$.
 This approximation is quite helpful if we like to go beyond three K\"ahler moduli. 
Thus, as in single modulus case in section \ref{sec:revi-analyt-study}, we may use the approximate potential (\ref{rewritten potential for Kahler}) to understand the statistical properties of the potential (\ref{effective potential for multi Kahler}) for the general multi-K\"ahler moduli cases studied in section \ref{sec:numer-analys-distr}.

\bibliographystyle{utphys}
\bibliography{C:/tex/papers/myrefs}

\end{document}